\documentclass[paper]{JHEP}
\usepackage{epsfig}
\usepackage{amssymb}
\def\beq{\begin{equation}}
\def\beqn{\begin{eqnarray}}
\def\eeq{\end{equation}}
\def\eeqn{\end{eqnarray}}
\def\abs#1{\left|#1\right|}

\def\FWeq#1{eq.~({\bf I}.#1)}
\def\FNWeq#1{eq.~({\bf II}.#1)}
\def\FKSeq#1{eq.~({\bf FKS}.#1)}

\def\HW{{\small HERWIG}}

\def\bk{\bar k}
\def\bl{\bar l}
\def\bp{\bar p}
\def\bs{\bar s}
\def\bt{\bar t}
\def\bu{\bar u}
\def\bd{\bar d}

\def\bx{\bar x}
\def\bbeta{\bar\beta}
\def\tbeta{\tilde\beta}
\def\dsb{d\bar\sigma}

\def\half{\frac{1}{2}}
\def\vep{\varepsilon}
\def\IN{{\small IN}}
\def\OUT{{\small OUT}}
\def\INm{{\sss IN}}
\def\OUTm{{\sss OUT}}

\newcommand\sss{\scriptscriptstyle\rm}

\newcommand\Delm{\Delta m_{12}}
\newcommand\Delmf{\left(\Delta m_{12}^2\right)^2}
\newcommand\Sigm{\Sigma m_{12}}
\newcommand\twototwo{2\to 2}
\newcommand\twotothree{2\to 3}

\newcommand\fo{f_1}
\newcommand\ft{f_2}

\newcommand\fsa{f_\alpha}
\newcommand\fot{f_{1,2}}

\newcommand\xMCB{\Big|_{\sss {\rm MC}}}
\newcommand\xMCBB{{\Bigg|_{\sss {\rm MC}}}}
\newcommand\clH{{\mathbb H}}
\newcommand\clS{{\mathbb S}}
\newcommand\EVprjmap{{\cal P}_{\clH\to\clS}}
\newcommand\EVprjmapi{{\cal P}_{\clH\to\clS}^{(\INm)}}
\newcommand\EVprjmapo{{\cal P}_{\clH\to\clS}^{(\OUTm)}}
\newcommand\bSigma{\overline{\Sigma}}
\newcommand\evBB{{\Bigg|_{\rm ev}}}
\newcommand\cntBB{\Bigg|_{\rm ct}}
\newcommand\Hone{(H_1)}
\newcommand\Htwo{(H_2)}
\newcommand\Itwo{{\cal I}_2}
\newcommand\Iqtwo{{\cal I}_{\tilde {2}}}

\newcommand\as{\alpha_{\sss S}}
\newcommand\gs{g_{\sss S}}
\newcommand\stepf{\Theta}
\newcommand\mydot{\!\cdot\!}
\newcommand\ub{\bar{u}}
\newcommand\db{\bar{d}}
\newcommand\bb{\bar{b}}

\newcommand\Sin{{\cal S}^{(\INm)}}
\newcommand\Sout{{\cal S}^{(\OUTm)}}
\newcommand\Sz{{\cal S}^{(0)}}
\newcommand\So{{\cal S}^{(1)}}
\newcommand\Szi{{\cal S}_i^{(0)}}
\newcommand\Soij{{\cal S}_{ij}^{(1)}}
\newcommand\Sztw{{\cal S}_2^{(0)}}
\newcommand\Szth{{\cal S}_3^{(0)}}

\newcommand\Sothtw{{\cal S}_{32}^{(1)}}

\newcommand\matr{{\cal M}^{(r)}}
\newcommand\matb{{\cal M}^{(b)}}

\newcommand\xic{\left(\frac{1}{\xi_i}\right)_c}
\newcommand\omyid{\left(\frac{1}{1-y_i}\right)_\delta}
\newcommand\opyid{\left(\frac{1}{1+y_i}\right)_\delta}
\newcommand\omyjd{\left(\frac{1}{1-y_j}\right)_\delta}
\newcommand\pt{p_{\sss T}}
\newcommand\kt{k_{\sss T}}
\newcommand\Et{E_{\sss T}}
\newcommand\ptrel{p_{\sss Trel}^{(h)}}
\newcommand\muR{\mu_{\sss R}}
\newcommand\muF{\mu_{\sss F}}

\preprint{
 Cavendish--HEP--05/24\hfill\\
 GEF--TH--12/2005\\
 ITP--UU--05/57\\
 NIKHEF/2005--026}
\title{Single-top production in MC@NLO}
\author{Stefano Frixione\\
  INFN, Sezione di Genova,
  Via Dodecaneso 33, 16146 Genova, Italy\\
  E-mail: \email{Stefano.Frixione@cern.ch}}
\author{Eric Laenen\\
  NIKHEF,  Kruislaan 409, 1098 SJ Amsterdam, The Netherlands and\\
  Institute for Theoretical Physics, Utrecht University, 
  Leuvenlaan 4, 3584 CE Utrecht, The Netherlands\\
  E-mail: \email{Eric.Laenen@nikhef.nl}}
\author{Patrick Motylinski\\
  NIKHEF,  Kruislaan 409, 1098 SJ Amsterdam, The Netherlands\\
  E-mail: \email{patrickm@nikhef.nl}}
\author{Bryan R.\ Webber\\
  Cavendish Laboratory, 
  J.J. Thompson Avenue, Cambridge CB3 0HE, U.K.\\
  E-mail: \email{webber@hep.phy.cam.ac.uk}}
\abstract{We match next-to-leading order QCD results for single-top 
hadroproduction with parton shower Monte Carlo simulations, according
to the prescription of the MC@NLO formalism. In this way, we achieve
the first practical implementation in MC@NLO of a process that has
both initial- and final-state collinear singularities. We show that 
no difficulties of principle arise from this complication, and present 
selected results relevant to the Tevatron.
}

\begin{document}

\section{Introduction}
Heavy flavour production at hadron colliders has been the subject
of extensive theoretical and experimental studies for more than
twenty years. The discovery of the top quark has offered 
an excellent opportunity to test QCD predictions much more reliably
than in the case of bottom or charm, thanks to the smaller value of $\as$
and the relatively minor impact of long-distance effects, the top
having no time to hadronize before decay. At present, all comparisons
between theory and data concern $t\bt$ pair production;
a crucial role in the satisfactory agreement between predictions 
and experimental results is played by the next-to-leading order (NLO) QCD 
corrections~\cite{Nason:1988xz,Beenakker:1989bq,Nason:1989zy,Beenakker:1991ma},
which enlarge the leading-order cross section by about 30\% at the Tevatron. 
A companion process to pair production is that in which a single top
quark is present in the final state. In such a case, a weak-interaction $Wtb$
vertex is involved, and thus the single-$t$ cross section is smaller than the
one for $t\bt$ (in spite of being favoured by phase space volume), which 
so far has prevented observation of such a production mechanism by
Tevatron experiments.  In terms of Standard Model physics, single-$t$
production is a direct probe of the weak interactions of the top, which in
fact constitutes the main interest of single-$t$ signals. Amongst other
things, this may lead to measurements that have not been performed so far,
namely of the CKM matrix element $V_{tb}$, and of the $b$ parton density.  
Single-$t$ production is in addition an important background for many 
searches for new physics, and can in general be seen as an effective way
to study new physics phenomena in the heavy sector.

For single-$t$ searches, or counting experiments in which single-$t$ 
is a background, it is crucial to have a reliable estimate of the
number of events expected, i.e. of the total rate. In this respect,
NLO results are mandatory, also in view of the fact that they allow
a sensible assessment of the size of unknown contributions of higher
orders. Calculations of fully-differential NLO single-$t$ cross sections 
have been performed in 
refs.~\cite{Harris:2002md,Sullivan:2004ie,Sullivan:2005ar,Zhu:2002uj}
and, including NLO top quark decay, in 
refs.~\cite{Campbell:2004ch,Cao:2004ky,Cao:2004ap,Cao:2005pq,Campbell:2005bb}.
On the other hand, in order to optimize acceptance cuts in an
experimental analysis, or to perform full detector simulations, one
needs realistic hadron-level events, which are obtained with Monte
Carlo event generators that incorporate the simulation of parton
showers and hadronization models. 

The complementary benefits of fixed-order computations and parton shower
simulations have been discussed at length in the literature, as well as the
advantages of combining them into a framework which would retain the strong
points of each of them. The MC@NLO 
approach~\cite{Frixione:2002ik,Frixione:2003ei} (we shall refer to these
papers as to {\bf I} and {\bf II} respectively hereafter) provides a
way of achieving this, by allowing one to match cross sections computed 
at NLO in QCD with an event generator. No modifications to the latter are
necessary, and therefore {\em existing} parton shower Monte Carlos can be used
for this purpose.

Although the MC@NLO formalism has been defined in full generality in 
{\bf I}, explicit implementation details have
been given there only for processes with no final-state QCD emissions
at the level of hard reactions. Such a case has been considered later 
in {\bf II}, with the implementation of $t\bar{t}$ and of $b\bar{b}$
production. In the context of MC@NLO, a {\em process-independent}
calculation is required for each type of soft and/or collinear singularity
which appears in the NLO real matrix elements. A quick inspection of
the processes implemented so far (see ref.~\cite{Frixione:2005gz}) should
convince the reader that the {\em only} singularity structure untreated
is the final-state collinear one. We shall deal with this singularity
in the present paper. It must be clear that, as for all of the other
singularities which have been studied previously, our formulation will
not depend on the fact that the specific single-$t$ production process
is considered here: in the derivation of the analytical formulae the nature of
the hard reaction is irrelevant (which is further evidenced by the fact that
the inclusion of single-$t$ production in MC@NLO relies significantly on
results obtained in {\bf I} and {\bf II}). What we achieve here is therefore,
besides the addition of an important process to the MC@NLO framework, the
capability of including other processes in MC@NLO without the need of
performing further analytical computations, notably those having final-state
(massless) partons at lowest order.

The paper is organized as follows: in sect.~\ref{sec:xsec} we discuss
single-$t$ production in the context of fixed-order computations and
Monte Carlo simulations. Sect.~\ref{sec:ME} reviews the status of the
matrix elements used in the present computation. We limit ourselves
here to implementing the $s$- and $t$-channel production mechanisms, 
and neglect spin correlations in production.
In sect.~\ref{sec:sub} we show that some changes can be made in the
subtraction formalism~\cite{Frixione:1995ms,Frixione:1997np} upon
which MC@NLO is based, which leave its analytical expression unaffected,
but improve its numerical stability. We then proceed to 
sect.~\ref{sec:MCtoNLO}, where we write down the approximate single-$t$
production cross sections generated by \HW, which enter the definition
of the MC subtraction terms needed for the matching with NLO results.
Implementation details of MC@NLO, concerning in particular the 
simultaneous presence of initial- and final-state collinear singularities,
are given in sect.~\ref{sec:MCatNLO}. We present results for single-$t$
production at the Tevatron in sect.~\ref{sec:res}; phenomenological studies,
including results for the LHC, will be the subject of a future paper. Finally,
conclusions and future prospects are reported in sect.~\ref{sec:concl}.
Some technical details are collected in the Appendices.

\section{Single-top cross sections\label{sec:xsec}}
Each process in MC@NLO is based on two main building blocks: a 
fully-exclusive NLO computation; and the knowledge of the so-called 
MC subtraction terms, which are closely related to the first
non-trivial order in the formal $\as$ expansion of the \HW\ Monte
Carlo result. We shall treat these two issues in turn.

\subsection{NLO computation\label{sec:NLO}}
Fully-exclusive observable predictions do not strictly exist in QCD: the theory
has finite resolution power, in the sense defined by the KLN
theorem. However, we can conventionally talk of fully-exclusive computations,
as those in which the cancellation of the infrared singularities is 
{\em formally} achieved analytically in an observable-independent manner, 
and the four-momenta of all of the final-state partons are available for 
defining the observables -- this does not violate the KLN theorem, since 
the formal cancellation mentioned above actually occurs only in the case
of infrared-safe observables. Fully-exclusive computations are crucial
for the matching of NLO cross sections with Monte Carlos, since the
latter need to know the four-momenta of all the particles involved in
the hard process in order to compute the initial conditions and the
various branching probabilities for the parton showers. Modern computations
of this kind are based on universal subtraction or slicing formalisms; we
shall discuss the one used within MC@NLO in sect.~\ref{sec:sub}. Before
doing that, we give some details specific to the matrix elements for
single-$t$ production.

\subsubsection{Matrix elements\label{sec:ME}}
\begin{figure}[htb]
  \begin{center}
      \epsfig{figure=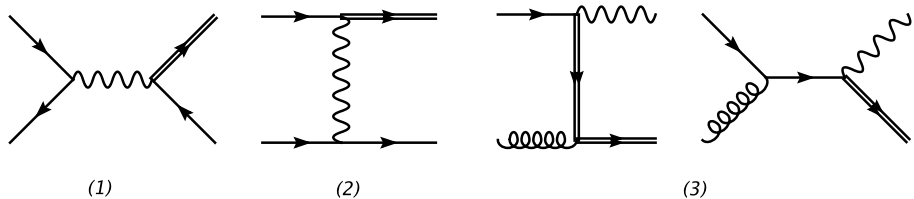,width=0.8\textwidth}
\caption{\label{fig:one} 
Leading order diagrams for single-$t$ production in the 
(1) $s$-channel, (2) $t$-channel and (3) $Wt$-mode. The $t$-quark
 line is doubled.}
  \end{center}
\end{figure}

The lowest-order parton level processes are customarily divided into
three classes that will also serve to categorize the NLO contributions. 
They are shown in fig.~\ref{fig:one}. In the first diagram the single 
top quark is produced in the annihilation process 
\begin{equation}
  \label{eq:1}
  u + \bd \rightarrow t + \bb\,,
\end{equation}
via a time-like $W$ boson, and is therefore called the
$s$-channel process. In the second diagram, the 
initial bottom quark is converted into a top quark via the 
exchange of a $W$-boson
\begin{equation}
  \label{eq:2}
  b + u \rightarrow t + d\,,
\end{equation}
and is therefore called the $t$-channel process.  The final two graphs
represent the $Wt$ process in which the top quark is produced in association
with a real $W$
\begin{equation}
  \label{eq:3}
  b + g \rightarrow W + t\,.
\end{equation}
The cross section for this process occurring at the Tevatron is very small and
we neglect it in this paper. For the LHC this process becomes non-negligible
however. Note that in reactions (\ref{eq:1}) and (\ref{eq:2}) we have only
listed the CKM-dominant combinations of quark flavours, but all CKM-allowed
combinations are included in this paper. Consistently, the $b$ quark is 
always assumed to be massless.

\begin{figure}[htb]
  \begin{center}
      \epsfig{figure=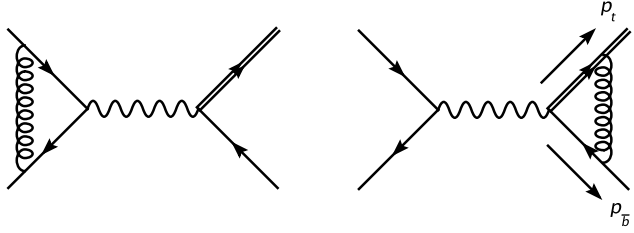,width=0.55\textwidth}
\caption{\label{fig:two} 
Virtual vertex corrections to $s$-channel single-$t$ production.}
  \end{center}
\end{figure}

In NLO one must include virtual and real corrections to the $s$- and
$t$-channel processes. The virtual corrections consist of vertex corrections
to diagrams {\sl (1)} and {\sl (2)} in fig.~\ref{fig:one}, together 
with the self-energy corrections to the $t$-quark line\footnote{Box graphs 
vanish since they involve a single colour matrix on a fermion line, 
i.e. a null trace.}. We shall not discuss these corrections in detail, nor 
give their explicit expressions, as these are already given in the
literature. To prepare a remark on single-antitop production further below,
we recall here that the vertex correction in the first diagram of
fig.~\ref{fig:two} is proportional to the lowest order vertex $\gamma^\mu
(1-\gamma_5)$ because only light quark lines are attached to it. If the top
quark line is attached as in the second diagram of fig.~\ref{fig:two}, a
second form factor appears at NLO, proportional to the difference $(p_t^\mu -
p_{\bar b}^\mu)/m_t$. A similar situation occurs in the $t$-channel.

Concerning the real-emission corrections, we categorize these processes by
the dominant CKM contributions, as follows
\begin{eqnarray}
u\db&\longrightarrow& t\bb g\,,
\label{eq:one}
\\
ub&\longrightarrow& tdg\,,
\label{eq:two}
\\
b\db&\longrightarrow& t\ub g\,,
\label{eq:three}
\\
ug&\longrightarrow& t\bb d\,,
\label{eq:four}
\\
\db g&\longrightarrow& t\bb\ub\,,
\label{eq:five}
\\
bg&\longrightarrow& td\ub\,.
\label{eq:six}
\end{eqnarray}
A rather detailed discussion of these processes, and how they
are assigned to $s$- and $t$-channel, can be found in the next section.
\begin{figure}[htb]
  \begin{center}
      \epsfig{figure=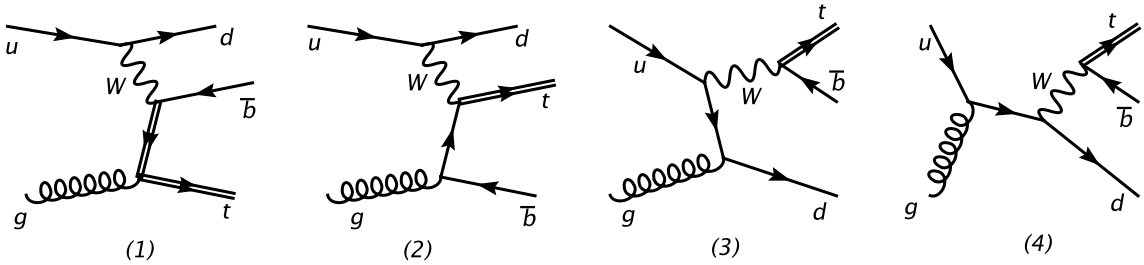,width=1.0\textwidth}
\caption{\label{fig:four} 
Diagrams contributing to $ug\to t\bb d$.
}
  \end{center}
\end{figure}
\begin{figure}[htb]
  \begin{center}
      \epsfig{figure=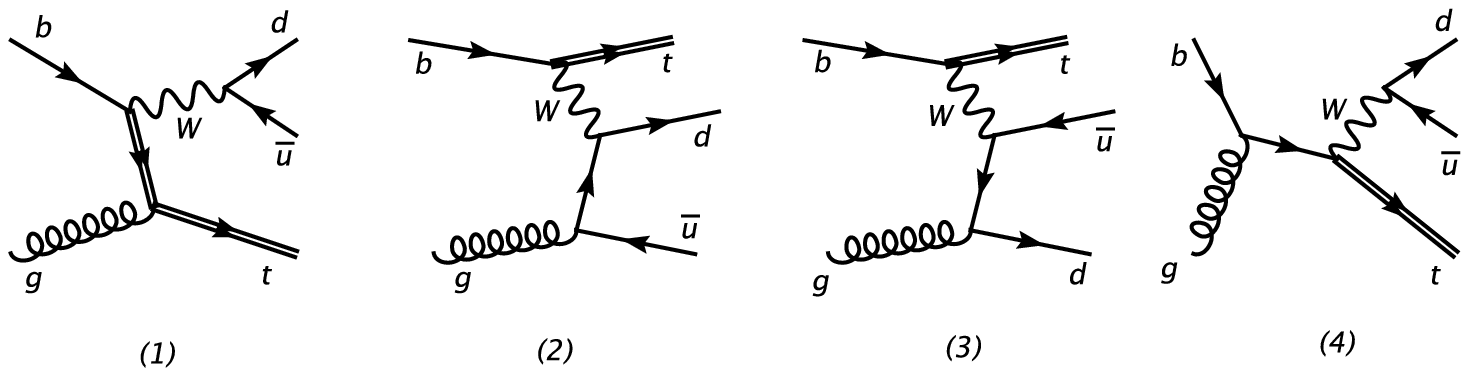,width=1.0\textwidth}
\caption{\label{fig:six} 
Diagrams contributing to $b g\to td\ub$.
}
  \end{center}
\end{figure}

The calculation of the single-$\bt$ cross section is perfectly
similar to that for the single-$t$ described above, after charge 
conjugation. It may be perhaps less apparent that the
second vertex form factor, mentioned above, proportional to 
$(p_{\bar t} - p_b)/m_t$ remains unchanged, since the quark propagators 
change the sign of the mass term. However the charge flow of the conjugated
amplitude is also reversed, resulting in an unchanged expression.

\subsubsection{Subtraction procedure\label{sec:sub}}
In order to implement a process in MC@NLO, its NLO cross section must be
computed according to the subtraction formalism presented in
refs.~\cite{Frixione:1995ms,Frixione:1997np} (denoted as FKS henceforth).
The basic idea in FKS is that of partitioning the phase space of the
final-state partons involved in real-emission contributions, in such a 
way that the resulting regions do not overlap, cover the whole phase space,
and each of them contains at most one collinear and one soft singularity. 
In each of these regions it is natural to select the one parton
(called the FKS parton here) with which the singularities are associated.
Denoting by $\matr$ the generic real matrix elements, this amounts to 
writing\footnote{The notation of refs.~\cite{Frixione:1995ms,Frixione:1997np}
has been slightly changed here in order to simplify the discussion. Functions
${\cal S}$ of the present paper play the same role as functions $\stepf$
in ref.~\cite{Frixione:1997np}.}
\begin{eqnarray}
1&=&\sum_i \Szi + \sum_{ij} \Soij\,,
\label{Sident}
\\
\matr&=&\sum_i \Szi\matr + \sum_{ij} \Soij\matr\,.
\label{FKSpart}
\end{eqnarray}
The FKS parton is labelled with $i$ in eqs.~(\ref{Sident}) 
and~(\ref{FKSpart}). The first term on the r.h.s. of eq.~(\ref{FKSpart}) gives
a divergent contribution (i.e., a contribution which has to be subtracted)
only in the infrared regions in which parton $i$ is soft and/or collinear to
one of the initial-state partons. Analogously, the only infrared regions in
which the second term on the r.h.s. of eq.~(\ref{FKSpart}) is divergent are
those in which parton $i$ is soft and/or collinear to final-state parton
$j$. More precisely, denoting by $p_\alpha$ and $k_\alpha$ the four-momenta of
the initial- and final-state particles respectively, we have
\begin{eqnarray}
&&\lim_{k^0_i\to 0}\left(\Szi+\sum_j \Soij\right)=1\,,
\label{Slimsf}
\\
&&\lim_{\vec{k}_i\parallel \vec{p}_1}\Szi=1\,,
\label{Slimcl1}
\\
&&\lim_{\vec{k}_i\parallel \vec{p}_2}\Szi=1\,,
\label{Slimcl2}
\\
&&\lim_{\vec{k}_i\parallel \vec{k}_j}\Soij=1\,,
\label{Slimclf}
\end{eqnarray}
while all the other infrared limits not explicitly listed above are 
zero\footnote{The superscripts $(0)$ and $(1)$ are legacy notation from 
ref.~\cite{Frixione:1995ms}, where these $S$-functions are related to
jet-finding algorithms, and the superscripts indicate the algorithm 
step at which a merging takes place.}.
Eqs.~(\ref{Slimsf})--(\ref{Slimclf}) are the {\em only} properties
of the ${\cal S}$ functions used in the analytical computations of
refs.~\cite{Frixione:1995ms,Frixione:1997np}; their actual functional
forms away from the infrared limits are only relevant to numerical
integrations. It should be stressed that all partons in the final state
may induce a divergence of the real matrix elements; to take this fact
into account, the role of FKS parton is given to each parton in turn,
which is formally expressed in eq.~(\ref{FKSpart}) by the sum over $i$ 
that appears on the r.h.s. there.

After the phase space of the final-state partons is effectively 
partitioned through eq.~(\ref{FKSpart}) into different infrared-singular
regions, FKS chooses a different phase-space {\em parametrization} 
in each of these regions. It must be clear that the phase space is
always the same, i.e. that relevant to the $n$ particles involved in
real-emission processes; the only difference between the various 
regions is in the choice of the integration variables which are left
after getting rid of the $\delta$ functions that appear in the
phase-space definition. The integration variables are chosen to perform 
the necessary analytical integrations in an easy way, and to 
facilitate importance sampling in numerical integrations. 
The key variables in the phase-space parametrization associated 
with $\Szi$ are the energy of parton $i$ (directly related to soft
singularities), and the angle between parton $i$ and one of the
initial-state partons (directly related to initial-state collinear
singularities). For $\Soij$, the energy of parton $i$ and the angle
between parton $i$ and parton $j$ (related to a final-state collinear
singularity) are chosen instead. Obviously, the indices $i$ and $j$ are 
dummy here (phase spaces are flavour blind), and therefore there are only two
independent functional forms for phase spaces in FKS, which loosely speaking
are relevant to initial- and to final-state emissions. More details, and
specific functional forms, are given in appendix~\ref{sec:MCsubt}.

After the partition of the phase space, achieved by means of $\Szi$ 
and $\Soij$, it is the matrix elements that determine whether a singularity 
actually occurs in a given region of such a partition. As a general rule,
one should choose the simplest possible forms for the ${\cal S}$ functions
that still allow subtraction of all singularities. Although this is 
by no means mandatory (a region without singularities will simply give
a finite contribution to the cross section), it is beneficial for 
well-behaved numerical computations. Since single-$t$ matrix elements
have a singularity structure much simpler than that of the matrix 
elements considered in refs.~\cite{Frixione:1995ms,Frixione:1997np}, the 
${\cal S}$ functions will also be simpler here. We also want to use the 
present process as a test case, and will define the ${\cal S}$'s 
as smooth functions of invariants, at variance with the original
formulation of refs.~\cite{Frixione:1995ms,Frixione:1997np}, in which
they have been expressed as products of $\Theta$ functions. 

We start by denoting the four-momenta entering an NLO tree-level 
single-$t$ production process as follows
\begin{equation}
\alpha(p_1)+\beta(p_2)\,\longrightarrow\,t(k_1)+\gamma(k_2)+\delta(k_3)\,,
\label{TLproc}
\end{equation}
where $\alpha$ and $\beta$ are the incoming partons from the left
($p_1^3>0$) and from the right ($p_2^3<0$) respectively; $\gamma$ and
$\delta$ denote light final-state partons. We shall use the following 
shorthand notation
\begin{equation}
(\alpha,\beta;t,\gamma,\delta)
\label{TLshort}
\end{equation}
for the momentum assignment of eq.~(\ref{TLproc}). 

We first consider process~(\ref{eq:one}); the treatment of 
processes~(\ref{eq:two}) and~(\ref{eq:three}) is 
identical\footnote{It is immediate to see that the procedure adopted here
to disentangle the singularities of $(\alpha,\beta;t,\gamma,\delta)$ works
identically for $(\beta,\alpha;t,\gamma,\delta)$.}. We assign momenta as 
follows:
\begin{equation}
(u,\db;t,\bb,g)\,.
\label{eq:oneFKS}
\end{equation}
By inspection of the relevant Feynman diagrams, we immediately conclude that
the only singularities are associated with the gluon: the final-state light
quark cannot give rise to a collinear divergence, being in all cases
connected to a $W$ boson. Therefore, for such processes the gluon will always
be the FKS parton and, according to the discussion given at the
beginning of this section, we can choose the ${\cal S}$ functions in such
a way that the only non-zero ones are $\Szth$ and $\Sothtw$. In particular, 
with the following forms
\begin{eqnarray}
\Szth&=&\frac{(k_3\mydot k_1)^a (k_3\mydot k_2)^a}
{(k_3\mydot k_1)^a (k_3\mydot k_2)^a + (k_3\mydot p_1)^a (k_3\mydot p_2)^a}\,,
\label{Sindef}
\\
\Sothtw&=&\frac{(k_3\mydot p_1)^a (k_3\mydot p_2)^a}
{(k_3\mydot k_1)^a (k_3\mydot k_2)^a + (k_3\mydot p_1)^a (k_3\mydot p_2)^a}\,,
\label{Soutdef}
\end{eqnarray}
equations~(\ref{Sident})--(\ref{Slimclf}) are fulfilled ($i\equiv 3$, the
gluon being the FKS parton). In eqs.~(\ref{Sindef}) and~(\ref{Soutdef}) 
$a$ is an {\em arbitrary} positive real number; the physical results 
will not depend on $a$, and their stability against the variation of $a$ 
will constitute a check of the correctness of our implementation. It is 
clear that the numerators of eqs.~(\ref{Sindef}) and~(\ref{Soutdef}) will
act as damping factors for final- and initial-state collinear singularities
respectively; the larger $a$, the stronger the damping. Formally, in the
$a\to\infty$ limit we could recover the $\stepf$-based implementation of the
${\cal S}$ functions of refs.~\cite{Frixione:1995ms,Frixione:1997np}. More
pragmatically, we shall use the freedom in the choice of $a$ to improve, 
if necessary, the numerical stability of the result, and will study its 
impact on the number of negative-weight events in MC@NLO, see 
section~\ref{sec:res}.

We now turn to the case of process~(\ref{eq:four}); the corresponding
Feynman diagrams are shown in fig.~\ref{fig:four}. There are only 
initial-state collinear singularities in this case, due to the
splittings $g\to b\bb$ (graph~2) and $g\to d\db$ (graph~3). On the
other hand, these two diagrams do not interfere: the former contributes
to the $t$-channel cross section, the latter to the $s$-channel one.
Since $s$- and $t$-channel contributions are integrated separately,
we are in the same situation as process~(\ref{eq:one}) (i.e., only
one parton can give singularities), except for the fact that no
final-state singularities are present in this case. Therefore,
we can set $\So=0$ here, which implies $\Sz=1$. It also implies 
that we are in the same situation as that treated in {\bf I}
(which also applies to many other processes implemented in MC@NLO).
This situation now naturally appears as a particular case of 
a more general implementation in which singularities are disentangled 
by means of ${\cal S}$ functions.

Since process~(\ref{eq:five}) is completely analogous to 
process~(\ref{eq:four}), we finally deal with process~(\ref{eq:six}), 
whose Feynman diagrams are shown in fig.~\ref{fig:six}. Of those,
graphs~1 and~4 contribute to the $Wt$ mode, which has not been
considered here and are therefore dropped, while graphs~2 and~3 contribute 
to the $t$-channel.  Graph~2 (graph~3) is singular when the $\ub$ ($d$) is 
emitted collinearly to the initial-state gluon; since the two diagrams do 
interfere, we disentangle the singularities by means of the ${\cal S}$ 
functions. We assign the momenta according to
\begin{equation}
(b,g;t,d,\ub)\,,
\label{eq:sixFKS}
\end{equation}
and we define
\begin{eqnarray}
\Sztw&=&\frac{(k_3\mydot p_1)^a (k_3\mydot p_2)^a}
{(k_2\mydot p_1)^a (k_2\mydot p_2)^a + (k_3\mydot p_1)^a (k_3\mydot p_2)^a}\,,
\label{Sz2}
\\
\Szth&=&\frac{(k_2\mydot p_1)^a (k_2\mydot p_2)^a}
{(k_2\mydot p_1)^a (k_2\mydot p_2)^a + (k_3\mydot p_1)^a (k_3\mydot p_2)^a}\,,
\label{Sz3}
\end{eqnarray}
which again fulfill equations~(\ref{Sident})--(\ref{Slimclf}). Although
the same arbitrary parameter $a$ as in eqs.~(\ref{Sindef}) 
and~(\ref{Soutdef}) has been used here, this is in fact not necessary;
we could introduce another free parameter, independent of $a$.

We conclude this section by stressing that the functional dependences
of the ${\cal S}$ functions given above are correlated with the momentum
assignments chosen for the corresponding subprocesses. For example,
eqs.~(\ref{Sz2}) and~(\ref{Sz3}) imply that the FKS parton will
have four-momentum $k_2$ and $k_3$ respectively. Clearly, the subtraction
formalism is independent of the particular labeling adopted for each
process. Therefore, through a relabeling we can always assign
four-momentum $k_3$ to the FKS parton. Such relabeling is a purely
formal trick to render manifest the local matching between NLO
matrix elements and MC subtraction terms.

As far as processes~(\ref{eq:one})--(\ref{eq:three}) are concerned, 
we pointed out before that only the gluon can play the role of FKS parton. 
Thus, the momentum assignment in eq.~(\ref{eq:oneFKS}) and the analogous 
ones
\begin{eqnarray}
(u,b;t,d,g)
\label{eq:twoFKS}
\\
(b,\db;t,\ub,g)
\label{eq:threeFKS}
\end{eqnarray}
are what we want; as a consequence, the ${\cal S}$ functions have still
the forms given in eqs.~(\ref{Sindef}) and~(\ref{Soutdef}). For the process 
in eq.~(\ref{eq:four}), see fig.~\ref{fig:four}, we noted that in the cases
of $s$- and $t$-channel contributions the singularities arise from
the splittings $g\to d\db$ and $g\to b\bb$ respectively. Therefore,
we assign momenta as follows
\begin{eqnarray}
&&(u,g;t,\bb,d)\phantom{aaaaa}s\!-\!{\rm channel},
\label{eq:fourFKSs}
\\
&&(u,g;t,d,\bb)\phantom{aaaaa}\,t\!-\!{\rm channel},
\label{eq:fourFKSt}
\end{eqnarray}
and analogously for process~(\ref{eq:five})
\begin{eqnarray}
&&(\db,g;t,\bb,\ub)\phantom{aaaaa}s\!-\!{\rm channel},
\label{eq:fiveFKSs}
\\
&&(\db,g;t,\ub,\bb)\phantom{aaaaa}\,t\!-\!{\rm channel}.
\label{eq:fiveFKSt}
\end{eqnarray}
Finally, owing to the fact that \mbox{$\Sztw\leftrightarrow\Szth$}
when \mbox{$k_2\leftrightarrow k_3$}, we write
\begin{eqnarray}
\matr(b,g;t,d,\ub)&=&
\Szth\matr(b,g;t,d,\ub)+\Sztw\matr(b,g;t,d,\ub)
\nonumber\\*
&=&\Szth\Big[\matr(b,g;t,d,\ub)+\matr(b,g;t,\ub,d)\Big].
\label{tricksix}
\end{eqnarray}
In other words, we shall assign the momenta in process~(\ref{eq:six})
in two different ways
\begin{eqnarray}
&&(b,g;t,d,\ub)\,,
\label{eq:sixFKS1}
\\
&&(b,g;t,\ub,d)\,,
\label{eq:sixFKS2}
\end{eqnarray}
and for each of them we multiply the corresponding matrix element 
times $\Szth$ given in eq.~(\ref{Sz2}); as shown in eq.~(\ref{tricksix}),
this is fully equivalent to eqs.~(\ref{eq:sixFKS})--(\ref{Sz3}).

As a final remark, we note that when keeping the same ordered
notation~(\ref{TLshort}) after charge conjugation the
treatment of real emission corrections to anti-top production
is perfectly analogous, and the inclusion of single-$\bt$ 
requires no extra work.

\subsection{MC cross sections expanded to NLO\label{sec:MCtoNLO}}
As discussed in {\bf I} and {\bf II}, in order to construct the
MC subtraction terms one needs the cross section 
obtained by keeping the first non-trivial order in the $\as$ expansion
of the parton shower Monte Carlo that will be matched with the NLO 
computation. As in the previous papers, the explicit results presented 
here are relevant to \HW. The most general form of the MC cross sections 
is given in \FNWeq{5.1}, which we rewrite as follows
\begin{equation}
d\sigma\xMCB=\sum_{\mu}\sum_{L}\sum_{l}
d\sigma_{\mu}^{(L,l)}\xMCB\,,
\label{MCatas}
\end{equation}
where the index $\mu$ generically indicates a collection of labels which 
unambiguously identify the $\twotothree$ partonic subprocess. The index
$L$ assumes the values $+$, $-$, $\fo$, and $\ft$ (the latter two
were denoted by $Q$ and $\bar{Q}$ in {\bf II}). The index $l$, which differs
per colour structure, assumes the values $q_i\mydot q_j$, where
$q_i$ and $q_j$ are the four-momenta of the colour partners
relevant to the emission considered; in this way, the shower
scale is 
\begin{equation}
E_0^2=|l|\equiv |q_i\mydot q_j|. 
\label{shwrscale}
\end{equation}
In {\bf II} we had $l=s,t,u$ (and $E_0^2=|l|/2$), but in the case 
of unequal masses this is not a convenient notation.  
Equations ({\bf II}.5.2)--({\bf II}.5.5) are also unchanged 
apart from notation
\begin{eqnarray}
d\sigma_{\mu}^{(+,l)}\xMCB &=&
\frac{1}{z_+^{(l)}}
f_a^{\Hone}(\bx_{1i}/z_+^{(l)})f_b^{\Htwo}(\bx_{2i})\,
d\hat{\sigma}_{\mu}^{(+,l)}\xMCB d\bx_{1i}\,d\bx_{2i}\,,
\label{eq:spl}
\\
d\sigma_{\mu}^{(-,l)}\xMCB &=&
\frac{1}{z_-^{(l)}}
f_a^{\Hone}(\bx_{1i})f_b^{\Htwo}(\bx_{2i}/z_-^{(l)})\,
d\hat{\sigma}_{\mu}^{(-,l)}\xMCB d\bx_{1i}\,d\bx_{2i}\,,
\label{eq:smn}
\\
d\sigma_{\mu}^{(\fo,l)}\xMCB &=&
f_a^{\Hone}(\bx_{1f})f_b^{\Htwo}(\bx_{2f})\,
d\hat{\sigma}_{\mu}^{(\fo,l)}\xMCB d\bx_{1f}\,d\bx_{2f}\,,
\label{eq:sfo}
\\
d\sigma_{\mu}^{(\ft,l)}\xMCB &=&
f_a^{\Hone}(\bx_{1f})f_b^{\Htwo}(\bx_{2f})\,
d\hat{\sigma}_{\mu}^{(\ft,l)}\xMCB d\bx_{1f}\,d\bx_{2f}\,,
\label{eq:sft}
\end{eqnarray}
where the flavours $a$ and $b$ of the incoming partons depend on the 
value of $\mu$. The short-distance cross sections that appear on the 
r.h.s. of eqs.~(\ref{eq:spl})--(\ref{eq:sft}) can be read from 
\FNWeq{5.6} and \FNWeq{5.8}
\begin{eqnarray}
d\hat\sigma_{\mu}^{(\pm,l)}\xMCB &=& \frac{\as}{2\pi}\,
\frac{d\xi_\pm^{(l)}}{\xi_\pm^{(l)}}dz_\pm^{(l)}
P^{(0)}_{a^\prime b^\prime}(z_\pm^{(l)})\,
\dsb_{\mu^\prime}
\stepf\left((z_\pm^{(l)})^2-\xi_\pm^{(l)}\right),
\label{eq:shin}
\\
d\hat\sigma_{\mu}^{(\fsa,l)}\xMCB &=& \frac{\as}{2\pi}\,
\frac{d\xi_{\fsa}^{(l)}}{\xi_{\fsa}^{(l)}}dz_{\fsa}^{(l)}
P^{(0)}_{a^\prime b^\prime}(z_{\fsa}^{(l)})\,\dsb_{\mu^\prime}
\stepf\left(1-\xi_{\fsa}^{(l)}\right)
\stepf\left(z_{\fsa}^{(l)}-
\frac{m_\alpha}{E_0\sqrt{\xi_{\fsa}^{(l)}}}\right),\phantom{aa}
\label{eq:shout}
\end{eqnarray}
where the $\stepf$'s account for \HW\ dead regions (see
sect.~4.3 of {\bf II}), and the flavours $a^\prime$, $b^\prime$,
and the values of $\mu^\prime$ can be determined by considering
the possible collinear splittings of the corresponding NLO
tree-level processes.
\begin{table}[htb]
\begin{center}
\begin{tabular}{|c||c|c|}
\hline
 & $(u,\db;t,\bb)$ & $(\db,u;t,\bb)$ \\\hline\hline
$(u,\db;t,\bb,g)$ &
  $\pm(\bp_1\mydot \bp_2);\fot(\bk_1\mydot \bk_2)$ & \\\hline
$(\db,u;t,\bb,g)$ &
  & $\pm(\bp_1\mydot \bp_2);\fot(\bk_1\mydot \bk_2)$ \\\hline\hline
$(u,g;t,\bb,d)$ &
  $-(\bp_1\mydot \bp_2)$                       & \\\hline
$(\db,g;t,\bb,\ub)$ & 
  & $-(\bp_1\mydot \bp_2)$                       \\\hline
$(g,u;t,\bb,d)$ & 
  & $+(\bp_1\mydot \bp_2)$                       \\\hline
$(g,\db;t,\bb,\ub)$ & 
  $+(\bp_1\mydot \bp_2)$                       & \\\hline
\end{tabular}
\end{center}
\caption{\label{tab:sch}
Short-distance contributions to MC subtraction terms, for the 
$s$-channel. The two columns correspond to the two possible 
$s$-channel Born cross sections, distinguished by the direction
of the incoming partons. For a given process, the entries 
show the emitting legs, and in round brackets the value of the shower 
scale $E_0$ (up to a sign), according to the possible colour flows.
}
\end{table}
\begin{table}[htb]
\begin{center}
\begin{tabular}{|c||c|c|}
\hline
 & $(b,u;t,d)$ & $(b,\db;t,\ub)$ \\\hline\hline
$(b,u;t,d,g)$ & 
  $+,\fo(\bp_1\mydot \bk_1)$; $-,\ft(\bp_2\mydot \bk_2)$ & \\\hline
$(b,\db;t,\ub,g)$ & 
  & $+,\fo(\bp_1\mydot \bk_1)$; $-,\ft(\bp_2\mydot \bk_2)$ \\\hline\hline
$(b,g;t,d,\ub)$ & 
  $-(\bp_2\mydot \bk_2)$                             & \\\hline
$(g,u;t,d,\bb)$ & 
  $+(\bp_1\mydot \bk_1)$                             & \\\hline
$(b,g;t,\ub,d)$ & 
  & $-(\bp_2\mydot \bk_2)$                             \\\hline
$(g,\db;t,\ub,\bb)$ & 
  & $+(\bp_1\mydot \bk_1)$                             \\\hline
\end{tabular}
\end{center}
\caption{\label{tab:tcho}
As in table~\ref{tab:sch}, for the $t$-channel, with $bu$- and 
$b\db$-initiated Born processes.
}
\end{table}
\begin{table}[htb]
\begin{center}
\begin{tabular}{|c||c|c|}
\hline
 & $(u,b;t,d)$ & $(\db,b;t,\ub)$ \\\hline\hline
$(u,b;t,d,g)$ & 
  $+,\ft(\bp_1\mydot \bk_2)$; $-,\fo(\bp_2\mydot \bk_1)$ & \\\hline
$(\db,b;t,\ub,g)$ & 
  & $+,\ft(\bp_1\mydot \bk_2)$; $-,\fo(\bp_2\mydot \bk_1)$ \\\hline\hline
$(u,g;t,d,\bb)$ & 
  $-(\bp_2\mydot \bk_1)$                             & \\\hline
$(g,b;t,d,\ub)$ & 
  $+(\bp_1\mydot \bk_2)$                             & \\\hline
$(\db,g;t,\ub,\bb)$ & 
  & $-(\bp_2\mydot \bk_1)$                             \\\hline
$(g,b;t,\ub,d)$ & 
  & $+(\bp_1\mydot \bk_2)$                             \\\hline
\end{tabular}
\end{center}
\caption{\label{tab:tcht}
As in table~\ref{tab:sch}, for the $t$-channel, with $ub$- and 
$\db b$-initiated Born processes.
}
\end{table}

As in {\bf II}, we use unbarred and barred symbols to denote quantities 
relevant to $\twotothree$ and $\twototwo$ processes respectively. The
momentum assignments for the former are given in eq.~(\ref{TLproc}),
while for the latter we use
\begin{equation}
\alpha^\prime(\bp_1)+\beta^\prime(\bp_2)\,\longrightarrow\,
t(\bk_1)+\gamma^\prime(\bk_2)\,,
\label{TLproc0}
\end{equation}
which we shorten in a way similar to eq.~(\ref{TLshort})
\begin{equation}
(\alpha^\prime,\beta^\prime;t,\gamma^\prime)\,.
\label{TLshort0}
\end{equation}
In MC cross sections expanded to NLO, $\twototwo$ momenta (entering
$\dsb$ on the r.h.s. of eqs.~(\ref{eq:shin}) and~(\ref{eq:shout}))
are obtained by means of a suitable projection of the corresponding
$\twotothree$ momenta. The exact form of the projection is
specific to the parton shower MC matched to the NLO computation,
and for \HW\ can be worked out as was done in {\bf II}. Here, we need 
to extend the formulae given in {\bf II}, in order to treat the case 
of final-state partons with unequal masses; explicit results are
given in appendix~\ref{sec:2to2kin}.

As far as flavour combinations are concerned, it is simpler to 
read eqs.~(\ref{eq:shin}) and~(\ref{eq:shout}) from right to left,
since this follows the logic which forms the basis of a parton shower.
The MC starts with a Born-level ($\twototwo$ for single-$t$ production)
process, and then lets each leg branch in all kinematically- and
flavour-allowed configurations possible. This implies that several
$\twotothree$ processes may be generated starting from a given $\twototwo$
process. We list all such processes explicitly in
tables~\ref{tab:sch}--\ref{tab:tcht}; the non-void entries give
non-zero contributions to eq.~(\ref{MCatas}). Thus, the index
$\mu$ that classifies the $\twotothree$ partonic processes can simply be
chosen so as to count all of the quantities that appear in the first 
columns of the tables. Parton legs where the branchings
occur are denoted by $+$, $-$, $\fo$, and $\ft$ ($\fo$ always coincides with
the top quark); given the parton that branches, and the hard subprocess, a 
colour connection is established which fixes the shower scale $E_0$ 
unambiguously. The shower scales to be used in eqs.~(\ref{eq:shin}) 
and~(\ref{eq:shout}) are equal to the absolute values of the dot products 
listed in tables~\ref{tab:sch}--\ref{tab:tcht}. We finally point out that 
the momentum assignments for the $\twotothree$ processes in the tables above
are the same as those adopted (after relabeling) in the context of the
pure NLO computation. This puts the NLO and MC cross sections on the
same footing from a notational viewpoint, which will be convenient
for the formal manipulations to be carried out in the next section.

\section{MC@NLO\label{sec:MCatNLO}}
The definition of the MC@NLO formalism is given in \FWeq{4.22}
or \FNWeq{2.1}:
\begin{eqnarray}
{\cal F}_{\mbox{\tiny MC@NLO}}&=&\sum_{\mu}\int dx_1 dx_2 d\phi_3\Bigg\{
{\cal F}_{\mbox{\tiny MC}}^{(3)}
\Bigg(\frac{d\bSigma_{\mu}^{(f)}}{d\phi_3}\evBB
-\frac{d\bSigma_{\mu}}{d\phi_3}\xMCBB\Bigg)
\nonumber \\*&& \phantom{a}
+{\cal F}_{\mbox{\tiny MC}}^{(2)}\Bigg[
-\frac{d\bSigma_{\mu}^{(f)}}{d\phi_3}\cntBB
+\frac{d\bSigma_{\mu}}{d\phi_3}\xMCBB
+\frac{1}{\Itwo}\Bigg(\frac{d\bSigma_{\mu}^{(b)}}{d\phi_2}
+\frac{d\bSigma_{\mu}^{(sv)}}{d\phi_2}\Bigg)
\nonumber \\*&& \phantom{a}
+\frac{1}{\Iqtwo}\Bigg(
\frac{d\bSigma_{\mu}^{(c+)}}{d\phi_2 dx}\evBB
+\frac{d\bSigma_{\mu}^{(c-)}}{d\phi_2 dx}\evBB\Bigg)
-\frac{1}{\Iqtwo}\Bigg(\frac{d\bSigma_{\mu}^{(c+)}}{d\phi_2 dx}\cntBB
+\frac{d\bSigma_{\mu}^{(c-)}}{d\phi_2 dx}\cntBB\Bigg)
\Bigg]\Bigg\}.\phantom{aaaa}
\label{eq:rMCatNLO}
\end{eqnarray}
There is a minor difference of notation with respect to \FNWeq{2.1}: the
indices for the sum over all partonic processes are denoted
in the present paper by $\mu$, consistently with what was done in
sect.~\ref{sec:MCtoNLO}. We refer the reader to {\bf I} and {\bf II} 
for all details relevant to the formalism. Single-$t$ production is
the first process implemented in MC@NLO in which {\em both} $\Szi$ and
$\Soij$ are non-zero for certain $i$ and $j$. This has direct implications
for eq.~(\ref{eq:rMCatNLO}), which we now discuss.

As shown in sect.~\ref{sec:MCtoNLO}, for a given choice of the index
$\mu$ which classifies the partonic processes the radiation pattern
in the MC cross section is determined by the values of the indices
$L$ and $l$. On the other hand, the possible radiation patterns at 
the NLO level are determined by the $\Szi$ and $\Soij$ functions.
Inspection of sect.~\ref{sec:sub} and of tables~\ref{tab:sch}--\ref{tab:tcht}
shows that, for a given $\mu$, there are {\em at most} one $\Sz$ and one 
$\So$ functions which are non-vanishing. Formally, this corresponds to 
defining two single-valued functions $i(\mu)$ and $j(\mu)$ such that 
$\Sz_{i(\mu)}$ and $\So_{i(\mu)j(\mu)}$ may be different from zero. 
This allows us to define the following quantities
\begin{equation}
\Sin_\mu={\cal S}_{i(\mu)}^{(0)}\,,\;\;\;\;\;\;
\Sout_\mu={\cal S}_{i(\mu)j(\mu)}^{(1)}\,,
\label{Siodef}
\end{equation}
where the labels \IN\ and \OUT\ are to remind us that 
$\Szi$ and $\Soij$ select kinematics configurations relevant to 
initial- and final-state collinear emissions respectively. Note that 
one of the functions in eq.~(\ref{Siodef}) may be still be vanishing 
(which is the case for ${\cal S}^{(1)}$ in processes~(\ref{eq:sixFKS1})
and~(\ref{eq:sixFKS2})), but there cannot be other non-vanishing ${\cal S}$ 
functions. In any case, from eqs.~(\ref{Siodef}) and~(\ref{Sident}) we obtain
\begin{equation}
\Sin_\mu+\Sout_\mu=1\;\;\;\;\;\;\;\;
\forall\,\mu\,.
\label{Sioid}
\end{equation}
Note that, since we have exploited relabeling invariance to assign the 
four-momentum $k_3$ always to the FKS parton, we have $i(\mu)\equiv 3$.
Furthermore, since the other massless final-state parton has four-momentum 
$k_2$, we also have $j(\mu)\equiv 2$. However, eq.~(\ref{Siodef}) holds
independently of relabeling invariance. Furthermore, it is clear that
an analogous equation must hold for any kind of hard reaction, and not 
only for single-$t$ production, as a direct consequence of the
definition of the FKS partition.

The term $\bSigma^{(f)}$ is proportional to the real-emission
matrix elements (see \FWeq{4.12} and \FWeq{4.13}); thus, according
to eq.~(\ref{FKSpart}), in eq.~(\ref{eq:rMCatNLO}) we understand
\begin{eqnarray}
\frac{d\bSigma_{\mu}^{(f)}}{d\phi_3}&=&
\Sin_\mu\frac{d\bSigma_{\mu}^{(f)}}{d\phi_3}+
\Sout_\mu\frac{d\bSigma_{\mu}^{(f)}}{d\phi_3}
\\*&=&
\Sin_\mu\frac{d\bSigma_{\mu}^{(f)}}{d\phi_3^{(\INm)}}+
\Sout_\mu\frac{d\bSigma_{\mu}^{(f)}}{d\phi_3^{(\OUTm)}}\,,
\label{Ssplit}
\end{eqnarray}
where we have introduced two different {\em parametrizations}
($d\phi_3^{(\INm)}$ and $d\phi_3^{(\OUTm)}$) of the three-body
phase space $d\phi_3$, analogously to what is done in FKS in the
context of pure NLO computations (see sect.~\ref{sec:sub}).
Their explicit forms, which are irrelevant in what follows, will
be given in app.~\ref{sec:MCsubt}.

In eq.~(\ref{eq:rMCatNLO}) each point $(x_1,x_2,\phi_3)$ corresponds 
to a $\twotothree$ kinematic configuration (called $\clH$). In previous
MC@NLO implementations, a definite $\twototwo$ configuration (called $\clS$)
was chosen given $(x_1,x_2,\phi_3)$, according to a mapping $\EVprjmap$
whose form is dictated by \HW. The definition of such an unique mapping
requires elaborate manipulations of the MC subtraction terms since,
as shown in \FNWeq{B.32} and \FNWeq{B.33}, initial- and final-state
emissions would naturally lead to the definitions of two different
mappings $\EVprjmapi$ and $\EVprjmapo$. Following the same arguments 
as in app.~B of {\bf II}, we could implement single-$t$ production
using an unique $\EVprjmap$; as discussed there, however, this may
degrade the numerical accuracy in the integration step and the unweighting
efficiency. Furthermore, the two mappings $\EVprjmapi$ and $\EVprjmapo$
are a perfect match to the FKS phase-space partition, which enters
eq.~(\ref{eq:rMCatNLO}) through eq.~(\ref{Ssplit}); the mapping $\EVprjmapi$ 
($\EVprjmapo$) can be naturally expressed in terms of the variables of 
$d\phi_3^{(\INm)}$ ($d\phi_3^{(\OUTm)}$). In practice, we replace
\FWeq{4.23} and \FWeq{4.24} with the following pairs of equations
\begin{eqnarray}
I_{\clH}^{(\INm)}&=&\sum_{\mu}\int dx_1 dx_2 d\phi_3^{(\INm)} 
\Bigg(\Sin_\mu\frac{d\bSigma_{\mu}^{(f)}}{d\phi_3^{(\INm)}}\evBB
-\frac{d\bSigma_{\mu}^{(L=\pm)}}{d\phi_3^{(\INm)}}\xMCBB\Bigg)\,,
\label{Iithreedef}
\\
I_{\clS}^{(\INm)}&=&\sum_{\mu}\int dx_1 dx_2 d\phi_3^{(\INm)} 
\Bigg[-\Sin_\mu\frac{d\bSigma_{\mu}^{(f)}}{d\phi_3^{(\INm)}}\cntBB
+\frac{d\bSigma_{\mu}^{(L=\pm)}}{d\phi_3^{(\INm)}}\xMCBB
\nonumber \\*&& \phantom{a}
+\frac{\Sin_\mu}{\Itwo}\Bigg(\frac{d\bSigma_{\mu}^{(b)}}{d\phi_2^{(\INm)}}
+\frac{d\bSigma_{\mu}^{(sv)}}{d\phi_2^{(\INm)}}\Bigg)
+\frac{1}{\Iqtwo}\Bigg(
\frac{d\bSigma_{\mu}^{(c+)}}{d\phi_2^{(\INm)} dx}\evBB
+\frac{d\bSigma_{\mu}^{(c-)}}{d\phi_2^{(\INm)} dx}\evBB\Bigg)
\nonumber \\*&& \phantom{a}
-\frac{1}{\Iqtwo}\Bigg(\frac{d\bSigma_{\mu}^{(c+)}}{d\phi_2^{(\INm)} dx}\cntBB
+\frac{d\bSigma_{\mu}^{(c-)}}{d\phi_2^{(\INm)} dx}\cntBB\Bigg)
\Bigg]\,,
\label{Iitwodef}
\end{eqnarray}
and
\begin{eqnarray}
I_{\clH}^{(\OUTm)}&=&\sum_{\mu}\int dx_1 dx_2 d\phi_3^{(\OUTm)} 
\Bigg(\Sout_\mu\frac{d\bSigma_{\mu}^{(f)}}{d\phi_3^{(\OUTm)}}\evBB
-\frac{d\bSigma_{\mu}^{(L=\fot)}}{d\phi_3^{(\OUTm)}}\xMCBB\Bigg)\,,
\label{Iothreedef}
\\
I_{\clS}^{(\OUTm)}&=&\sum_{\mu}\int dx_1 dx_2 d\phi_3^{(\OUTm)} 
\Bigg[-\Sout_\mu\frac{d\bSigma_{\mu}^{(f)}}{d\phi_3^{(\OUTm)}}\cntBB
+\frac{d\bSigma_{\mu}^{(L=\fot)}}{d\phi_3^{(\OUTm)}}\xMCBB
\nonumber \\*&& \phantom{a\int dx_1 dx_2 d\phi_3^{(\OUTm)}}
+\frac{\Sout_\mu}{\Itwo}\Bigg(\frac{d\bSigma_{\mu}^{(b)}}{d\phi_2^{(\OUTm)}}
+\frac{d\bSigma_{\mu}^{(sv)}}{d\phi_2^{(\OUTm)}}\Bigg)\Bigg]\,,
\label{Iotwodef}
\end{eqnarray}
in such a way that the total rate is now 
\begin{equation}
\sigma_{tot}=I_{\clS}^{(\INm)}+I_{\clH}^{(\INm)}+
             I_{\clS}^{(\OUTm)}+I_{\clH}^{(\OUTm)}\,,
\label{stotII}
\end{equation}
which replaces \FWeq{4.25}. It should be clear that 
eqs.~(\ref{Iithreedef})--(\ref{Iotwodef}) are a direct consequence 
of the definition of MC@NLO: in fact, eq.~(\ref{eq:rMCatNLO}) 
is recovered by inserting ${\cal F}_{\mbox{\tiny MC}}^{(3)}$ on 
the r.h.s. of eqs.~(\ref{Iithreedef}) and~(\ref{Iothreedef}), and 
${\cal F}_{\mbox{\tiny MC}}^{(2)}$ on the r.h.s. of eqs.~(\ref{Iitwodef}) 
and~(\ref{Iotwodef}). 
We note that the Born ($\bSigma^{(b)}$) and the soft-virtual
($\bSigma^{(sv)}$) contributions have been manipulated similarly to 
what was done for the real-emission contribution in eq.~(\ref{Ssplit});
although strictly speaking this is not necessary, since these terms
are finite and therefore not involved in any subtraction, it helps
to improve the numerical evaluations of the integrals $I_{\clS}^{(\INm)}$
and $I_{\clS}^{(\OUTm)}$. On the other hand, the remainders of the 
initial-state collinear subtraction ($\bSigma^{(c\pm)}$) only appear
in $I_{\clS}^{(\INm)}$, since $I_{\clS}^{(\OUTm)}$ is associated with
final-state emissions. We have also introduced two two-body phase-space
parametrizations $d\phi_2^{(\INm)}$ and $d\phi_2^{(\OUTm)}$, which are
analogous to their three-body counterparts. Finally, we have introduced 
the notation
\begin{equation}
\bSigma_{\mu}\xMCB=\bSigma_{\mu}^{(L=\pm)}\xMCB+
\bSigma_{\mu}^{(L=\fot)}\xMCB\,,
\label{MCssplit}
\end{equation}
which is the analogue of eq.~(\ref{Ssplit}) for MC subtraction terms.
The first and second terms on the r.h.s. of eq.~(\ref{MCssplit})
receive contributions from eq.~(\ref{eq:shin}) and~(\ref{eq:shout})
respectively (i.e. from initial-state and final-state branchings).
Taking into account the properties of the MC subtraction terms
(see app.~\ref{sec:MCsubt}), this implies that 
eqs.~(\ref{Iithreedef})--(\ref{Iotwodef}) are finite; in fact,
final-state singularities of real-emission matrix elements and
their corresponding counterterms are removed in eqs.~(\ref{Iithreedef})
and~(\ref{Iitwodef}) by $\Sin$, while initial-state singularities are 
removed in eqs.~(\ref{Iothreedef}) and~(\ref{Iotwodef}) by $\Sout$. 
Thus, the same procedure as in sect.~4.5 of {\bf I} can be used in 
order to generate the hard events that are given to the parton
shower as initial conditions.

As a concluding remark, we point out that the reason why the 
subtraction formalism of refs.~\cite{Frixione:1995ms,Frixione:1997np} 
appears to be particularly well suited for MC@NLO implementations
can be read from eqs.~(\ref{Iithreedef})--(\ref{Iotwodef}).
The partition of the phase space into collinear-like singular
regions gives the FKS parton the same role as the softest parton
emitted by an MC in the first branching after the generation of the hard
process. Since as explained in {\bf I} and {\bf II} the first branching
is the only one that matters for matching the MC with an NLO computation,
the FKS parton and the softest parton emerging from the first branching
in the shower are naturally paired in the definition of MC@NLO. Apart from
guaranteeing the local cancellation of IR singularities, such pairing also
allows a good control on the numerical stability of the result. It is also
important to recall that, in each of the IR singular regions defined by the
FKS partition, there are no unnecessary NLO subtractions: the only counterterm
contributing to the result is that relevant to the real matrix element
singularity present in that given region. This fact is very beneficial in
reducing the number of negative-weight events.

\section{Results\label{sec:res}}
In this section we present sample results for single-$t$ production 
at the Tevatron with $\sqrt{S}=1.96$ TeV. We limit ourselves 
here to comparing MC@NLO predictions with those obtained with \HW\
and with an NLO code we have written according to the subtraction
method of refs.~\cite{Frixione:1995ms,Frixione:1997np}, as discussed
in sect.~\ref{sec:sub}. As a preliminary step, we have checked that
our NLO results (with $\muR=\muF=m_t$) for the total rate and various
$t$ and $\bar{t}$ distributions are in excellent agreement with those of 
MCFM~\cite{Campbell:2004ch}. All of the predictions given in 
this section have been obtained by using the MRST2002 default PDF 
set~\cite{Martin:2002aw}, and by setting $m_t=178$~GeV, which result
in total rates equal to 1.045~pb and 0.406~pb for $t$- and
$s$-channel respectively. We have rescaled \HW\ results to the 
NLO cross section, since we are only interested in the comparison 
of shapes in the case of standard MC's. Also, we have only considered 
here \HW\ results for the $t$-channel contribution; studies that 
also involve the $s$-channel will be shown 
in a forthcoming paper. All the MC@NLO and \HW\ results (but not, 
of course, the NLO ones) include the hadronization of the partons 
in the final state; furthermore, we forced the $W$ emerging from
the top decay to decay into a pair of leptons. In order to reduce
as much as possible the statistical errors, we have generated 
$5\cdot 10^5$ events for each MC@NLO and \HW\ run\footnote{Clearly,
we are not suggesting to collect an integrated luminosity of 
${\cal O}(1)$~ab$^{-1}$ at the Tevatron. Here, we simply 
aim to expose the features of the two MC simulations with some precision.}. 
Finally, we stress that all of the fixed-order predictions 
presented here will be denoted as having NLO accuracy, even in the case 
of observables which, in the sense of perturbation theory, are effectively of
leading order (see e.g. $\pt^{(tj)}$ below); this is consistent with the
terminology one {\em needs} to adopt in the context of MC@NLO (sect.~2.3 of
{\bf I}).

\begin{figure}[htb]
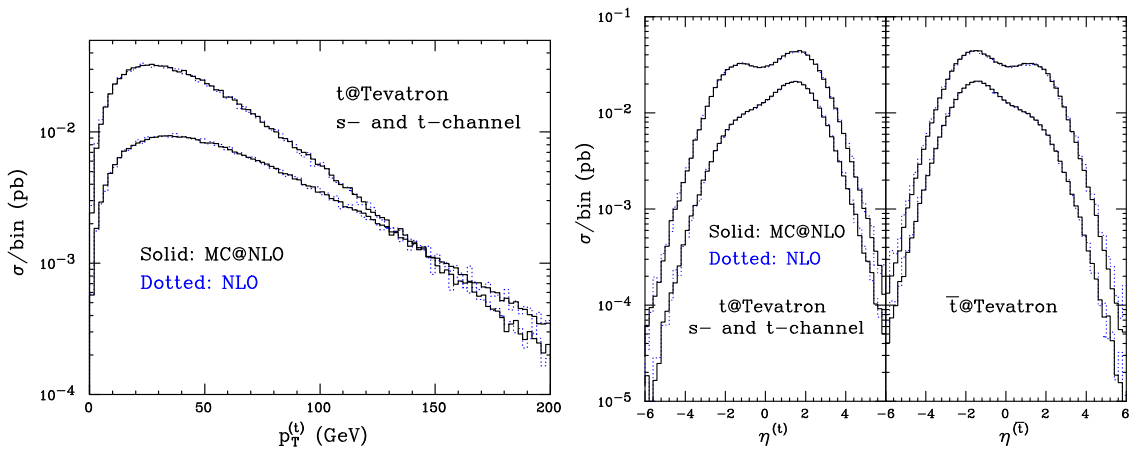

  \begin{center}
    \epsfig{figure=ptt_tsch_tev.eps,width=0.49\textwidth}
    \epsfig{figure=etattb_tsch_tev.eps,width=0.49\textwidth}
\caption{\label{fig:top_ptandeta} 
Comparison of MC@NLO (solid) and NLO (dotted) results. Left pane: top $\pt$, 
for $t$-channel (higher peak) and $s$-channel (lower peak) contributions. 
We have checked that $\pt^{(t)}=\pt^{(\bar{t})}$.
Right pane: top (left) and antitop (right) $\eta$, for $t$-channel 
(higher curves) and $s$-channel (lower curves) contributions.
}
  \end{center}
\end{figure}
We start by considering the transverse momentum and pseudorapidity
of the top and antitop (see fig.~\ref{fig:top_ptandeta}). We expect
the impact of the momentum reshuffling that takes place during the 
hadronization phase in MC@NLO to be negligible on such observables.
We also expect these observables, being sufficiently inclusive, 
to be reliably predicted by pure-NLO computations. As we see from the figure,
the good agreement between MC@NLO and NLO confirms our expectations,
and suggests that NNLO effects should be small. We have found that the
\HW\ results are extremely close to the MC@NLO ones, and for this reason
are not shown on the plots. As for all other processes previously
studied, we have observed a much-improved behaviour from the numerical
point of view when going from NLO to MC@NLO predictions, which is
due to the fact that in MC@NLO all cancellations between
large numbers occur at the level of short-distance cross sections,
rather than in histograms as in the case of NLO computations.
It is reassuring to see that this property holds true also for
single-$t$ production, which is the most involved process treated so 
far because of the simultaneous presence of initial- and final-state 
collinear singularities.

We now discuss the properties of a few jet observables. For the sake
of clarity, we limit ourselves in this discussion to 
considering $t$-channel top events. We reconstruct
the jets by means of the $\kt$-clustering algorithm~\cite{Catani:1993hr},
with $d_{cut}=100$~GeV$^2$. We include in the clustering procedure all
final-state stable hadrons\footnote{For the sake of simplicity, we
force $\pi^0$'s and all lowest-lying $b$-flavoured states to be stable
in \HW.} and photons. After the jets are reconstructed, we throw 
away the one that contains the $b$-flavoured hadron whose parent parton 
is the $b$ quark emerging from top decay, and order the remaining ones in
transverse energy, i.e. the hardest jet is the one with the largest $\Et$.
\begin{figure}[htb]
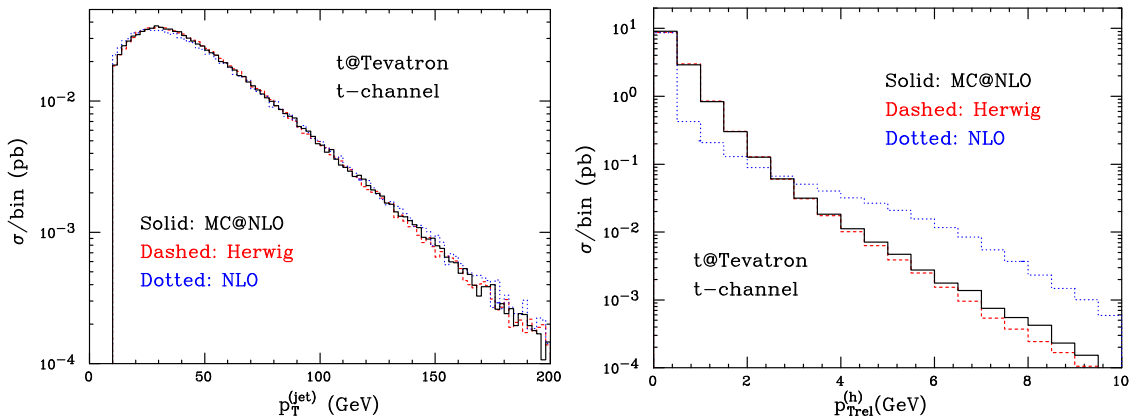

  \begin{center}
    \epsfig{figure=ptj1_tch_tev.eps,width=0.49\textwidth}
    \epsfig{figure=ptrelj1_tch_tev.eps,width=0.49\textwidth}
\caption{\label{fig:jsi} 
MC@NLO (solid), \HW\ (dashed), and NLO (dotted) results, for the
$\pt$ of the hardest jet (left pane), and the $\pt$ relative to the
axis of the hardest jet of those hadrons or partons in that jet
(right pane).
}
  \end{center}
\end{figure}

We recall that we do not let the top decay in our pure-NLO computation.
Also, we expect that some of the partons resulting from the radiation
by the $b$ quark emerging from the top decay in MC@NLO and \HW\ will
hadronize into hadrons that are not clustered into the $b$-jet
which we throw away. Furthermore, some extra radiation will occur from 
the top line due to showering, which is not included in the NLO computation. 
Finally, those jets obtained with MC@NLO and \HW\ are at the hadron level,
while those obtained with the NLO computation are at the parton level.

In spite of these differences, there is a good agreement between
MC@NLO, \HW, and NLO for the $\pt$ of the hardest jet, shown in the
left pane of fig.~\ref{fig:jsi}. This observable is sufficiently
inclusive for this to happen, and the small differences between
MC@NLO and NLO at small $\pt$ are mainly due to the hadronization
phase. On the other hand, the internal structure of the jet is
very different in MC@NLO and \HW\ from that resulting from
the NLO computation. In the right pane of fig.~\ref{fig:jsi} we
present the transverse momentum, relative to the axis of the jet,
of all of the hadrons or partons clustered into the jet itself.
At the NLO, the jet often coincides with a single parton, hence 
the sharp peak at $\ptrel=0$. Such a peak is much less pronounced
in the case of the MC's, since in those cases the jet almost never 
coincides with a single hadron. On the other hand, at large $\ptrel$ 
the MC results are smaller than the NLO one: this must be so, since in
the final states obtained with MC simulations it is likely that a 
large-$\ptrel$ hadron will be clustered into another jet. This is much
less probable at the NLO, simply because the number of jets there is limited
to two. It is also interesting to observe that, although very small, the
effect of the hard emissions due to the NLO real matrix elements is visible in
the tail of the $\ptrel$ distribution, the MC@NLO result being slightly
harder than the \HW\ one.

\begin{figure}[htb]
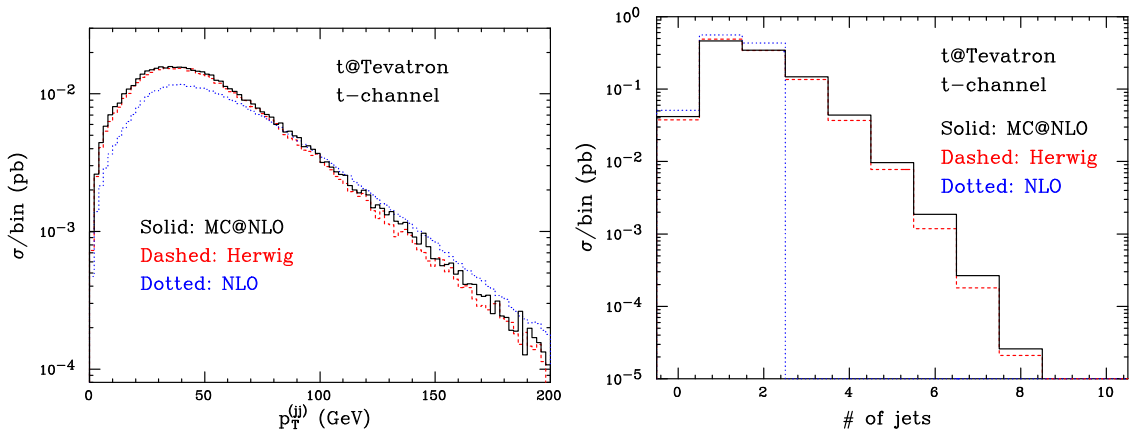

  \begin{center}
    \epsfig{figure=ptjj_tch_tev.eps,width=0.49\textwidth}
    \epsfig{figure=njet_tch_tev.eps,width=0.49\textwidth}
\caption{\label{fig:jjcorr} 
As in fig.~\ref{fig:jsi}, for the $\pt$ of the two-hardest-jet pair 
(left pane), and for the number of jets (right pane).
}
  \end{center}
\end{figure}
The differences between the topologies of the final states emerging from 
NLO computations and MC@NLO and \HW\ simulations are clearly visible when we 
consider observables less inclusive than the $\pt$ of the hardest jet. In
the left pane of fig.~\ref{fig:jjcorr} we plot the $\pt$ of the
pair of the two hardest jets. As is clear from the fact that MC@NLO
and \HW\ have very similar shapes, which are different from the NLO
one, the real matrix elements play a minor role here compared to the 
multiple emissions of the shower. The effects of the real matrix elements
are more clearly visible in the tail of the distribution in the number of 
final-state jets (right pane of fig.~\ref{fig:jjcorr}), with MC@NLO 
predicting more events with more than two jets compared to \HW.

\begin{figure}[htb]
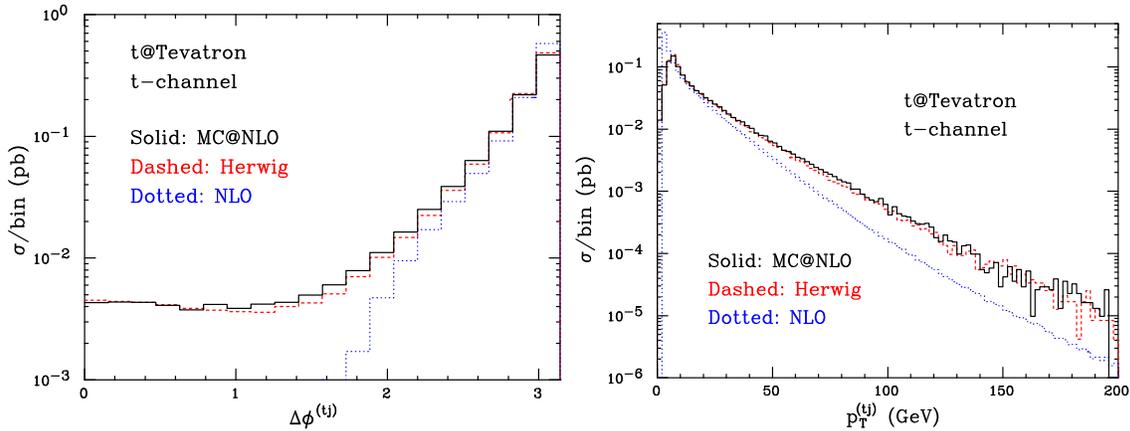

  \begin{center}
    \epsfig{figure=dftj1_tch_tev.eps,width=0.49\textwidth}
    \epsfig{figure=pttj1_tch_tev.eps,width=0.49\textwidth}
\caption{\label{fig:tjcorr} 
As in fig.~\ref{fig:jsi}, for the azimuthal difference between (left pane), 
and the $\pt$ of (right pane) the top-hardest jet pair.
}
  \end{center}
\end{figure}
It is also interesting to observe that shower effects dominate over
matrix element ones for top-hardest jet correlations, two of which
we present in fig.~\ref{fig:tjcorr}. We stress again here that we
did not make any systematic attempt to exclude from the jet clustering
the partons radiated by the top and its decay products, which would
allow a closer matching between MC's and NLO results for these
correlations. This is very clearly shown by the left pane of
fig.~\ref{fig:tjcorr}, which presents the difference in azimuth
between the top and the hardest jet. While the NLO prediction is
zero for $\Delta\phi^{(tj)}<\pi/2$ for kinematics reasons (there is
nothing in this region for the top-hardest jet pair to recoil
against), MC@NLO and \HW\ feature a long tail which extends down
to $\Delta\phi^{(tj)}=0$. This is in part due to the fact 
that the top tends to have a much larger longitudinal 
than transverse momentum component. Thus, it is relatively easy for 
a parton, radiated by the top quark shower, to change the top transverse
momentum by a sizable amount. The $\Delta\phi^{(tj)}=0$ tail is mainly
populated by such low-$\pt^{(t)}$ events. In the right pane of
fig.~\ref{fig:tjcorr} we present the $\pt$ of the top-hardest jet
pair. At the NLO level, only $\twotothree$ processes can contribute 
to the region $\pt^{(tj)}\ne 0$, in the configurations in which the 
two final-state massless partons are not combined into a single jet;
for this to happen, the two partons must be well separated. Clearly,
such configurations imply the presence of a very off-shell intermediate
particle, and are thus disfavoured by matrix elements: the $\pt^{(tj)}$
distribution is steeply falling. In MC@NLO and \HW, $\twotothree$
configurations result from a $\twototwo$ hard process followed
by a parton branching\footnote{In MC@NLO, there are also $\twotothree$
hard processes, whose matrix elements are the same as those of the
NLO computation.}. Since the branching is collinear in nature, the
probability of getting two well-separated partons is even smaller
than in NLO computations. However, the shower usually does not stop
after the first branching. Furthermore, all strongly-interacting
particles, including the top and the $b$ emerging from the top
decay, can radiate. This smears very effectively the final-state
momenta; we have verified that, in the large-$\pt^{(tj)}$ region, 
the hardest jet may retain a fraction of the parent parton momentum 
as small as 50\%. This creates an imbalance between the top and the
hardest-jet $\pt$ which results in the much harder $\pt^{(tj)}$ tails 
in MC@NLO and \HW\ relative to the NLO result. It should be stressed
that such an effect is magnified by the steepness of the $\pt^{(tj)}$ 
distribution. In terms of the total number of events, this is still
a marginal phenomenon, which gives a negligible contribution
to observables such as the inclusive $\pt$ of the hardest jet.
We conclude by observing again that the real matrix elements contributions
are small but visible in the differences between MC@NLO and \HW\ in
the intermediate $\Delta\phi^{(tj)}$ and large-$\pt^{(tj)}$ regions.

\begin{figure}[htb]
\begin{center}
\epsfig{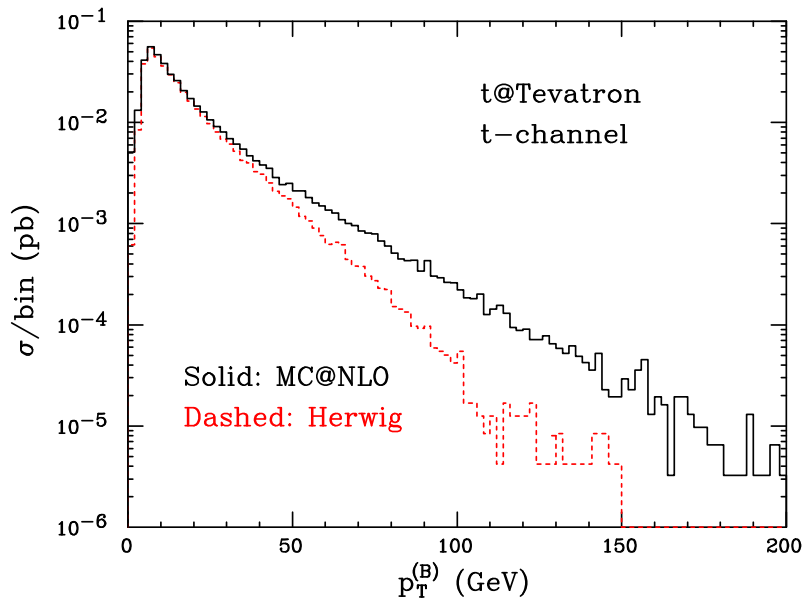}
\end{center}
\caption{\label{fig:ptb}
MC@NLO (solid) and \HW\ (dashed) results for the $\pt$ of the $b$-flavoured
hadrons (except those from top decay).
}
\end{figure}
The results presented so far have shown little or no difference between
MC@NLO and \HW\ results as far as shapes are concerned. Although larger
differences could be seen by imposing hard transverse momentum cuts,
the fact remains that at the Tevatron the phase-space for hard
radiation is fairly limited. There are, however, observables that
are particularly sensitive to real matrix element effects, such as the 
transverse momentum of the $b$-flavoured hadrons\footnote{$b$-flavoured 
hadrons from top decay are not included in this plot.}, which we
present in fig.~\ref{fig:ptb}. This is because in $t$-channel matrix
elements a $b$ quark is almost always present in the initial state
(up to CKM-suppressed contributions). This results in a final-state
$b$-flavoured hadron which, in the case of \HW, acquires its transverse
momentum entirely through the backward evolution in the shower mechanism. 
Such a mechanism is also present in MC@NLO, but there are also NLO real matrix
elements in which a $b$ quark has a large $\pt$, which is inherited by the
resulting $b$-flavoured hadron, and which explains the difference in the
large-$\pt^{(B)}$ tail between MC@NLO and \HW\footnote{For technical reasons,
fig.~\ref{fig:ptb} has been obtained by imposing $\abs{y^{(B)}}<3$. This cut
has no impact for $\pt^{(B)}>10$~GeV.}.

We conclude this section by mentioning the fact that we observe no 
dependence (within the statistical accuracy of the runs we performed)
of the physical results upon the unphysical parameters which enter the 
NLO subtraction formalism, such as the subtraction parameters introduced
in refs.~\cite{Frixione:1995ms,Frixione:1997np}, or the exponent
$a$ introduced in eqs.~(\ref{Sindef}) and~(\ref{Soutdef}). This
constitutes a test of the correctness of our implementation, 
since NLO results based on subtraction techniques are by
construction independent of these parameters. Similarly,
no dependence has been found on the parameters $\alpha$ and $\beta$
introduced in \FWeq{A.86} and \FWeq{A.87} which control the behaviour
of the MC subtraction terms in the soft limit, if they are restricted
to their natural ranges ($\alpha={\cal O}(1)$, $\beta={\cal O}(0.1)$).
This is as expected, since variation of these parameters gives only
power-suppressed effects.
On the other hand, all of the above parameters do affect the number of 
negative-weight events, and their tuning can be used to limit the presence 
of such events (whose fraction is equal to about 15\% in the results 
presented here). The parameter $a$ has only a limited impact on the
number of negative weights (which change by about 1\% for $1\le a\le 4$),
and its choice is mainly due to considerations of stability of the numerical
integration, with best results for $a=2$. In general, the accuracy of
the predictions obtained with values of $a$ larger than 2 (slowly) 
decreases with increasing $a$. Since the limit $a\to\infty$ corresponds 
to the $\stepf$-based implementation of the subtraction formalism, this 
indirectly proves that the implementation introduced in this paper is
more convenient from the numerical point of view.

\section{Conclusions\label{sec:concl}}
In this paper we have considered single-top hadroproduction in the
context of the MC@NLO approach. This case is, apart from its 
phenomenological relevance, also interesting from the technical point
of view, since it features both initial- and final-state collinear 
singularities, and thus has a radiation pattern different from
that of all of the processes so far included in MC@NLO.

We have shown that this is not a difficulty of principle, since
the MC@NLO formalism is unchanged with respect to its definition 
given in ref.~\cite{Frixione:2002ik}, but it entails a more involved
procedure in the generation of the hard events that are given to the
parton shower as initial conditions. Because this procedure is not 
specific to single-top hadroproduction, and since we have now
treated all possible radiation patterns in MC@NLO, we are now in
a position to include any new process, such as jet production, without 
the need of performing further analytical computations.

As in previous cases, our computation is based on the universal subtraction
formalism of refs.~\cite{Frixione:1995ms,Frixione:1997np}. We have used
single-top hadroproduction as a test case, to explore an implementation 
of the subtraction different from that of the original papers. The
partition of the phase space is now achieved by means of smooth functions
of invariants, rather than with $\stepf$ functions as was done previously.
This does not entail any change in the analytical formulae, but
helps to improve the behaviour of the numerical computations. There 
is also a conceptual difference, namely that the infrared singularities
are now disentangled by means of damping factors, rather than by
non-overlapping regions defined by the phase-space partition.
This in turn may lead to the possibility of implementing alternative 
subtraction schemes, although new analytical computations would be required 
in such a case.

We have not explored in this paper the phenomenological implications
of our work, since we limited ourselves to checking that all of the
observables we have considered show the expected behaviour in regions
where the NLO computations or the MC simulations should be most reliable.
We postpone the phenomenological studies, as well as the implementation of 
the $Wt$ mode and spin correlations, to a forthcoming paper.

\acknowledgments

We would like to thank the CERN TH division for 
hospitality during the completion of this work. The work of E.L. and P.M.  
is supported by the Netherlands Foundation for Fundamental Research of
Matter (FOM) and the National Organization for Scientific Research (NWO);
that of B.W. is supported in part by the UK Particle Physics and Astronomy 
Research Council. The Feynman diagrams in the paper have been drawn using 
the \textsf{Jaxodraw} package \cite{Binosi:2003yf}.

\appendix

\section{Kinematics\label{sec:2to2kin}}
In this section, we generalize the results of sect.~4 of {\bf II}
by considering the case of two final-state partons with unequal
masses. Consistently with {\bf II}, we use unbarred and barred 
symbols to denote quantities relevant to $\twotothree$ and $\twototwo$ 
processes respectively (see e.g. eqs.~(\ref{TLproc}) and~(\ref{TLproc0})
for four-momentum assignments). Although in single-$t$ production 
one of the final-state partons in $\twototwo$ processes is massless,
we shall derive our results in the most general case
\begin{equation}
\bk_1^2=m_1^2\,,\;\;\;\;\;\;
\bk_2^2=m_2^2\,.
\end{equation}
We start by defining the $\twototwo$ reduced invariants as follows
\begin{equation}
\bs_L = 2\bp_1\cdot\bp_2\,,\;\;\;\;\;\;
\bt_L =-2\bp_1\cdot\bk_1\,,\;\;\;\;\;\;
\bu_L =-2\bp_1\cdot\bk_2\,,
\label{eq:dot}
\end{equation}
with $L=+,-,\fo,\ft$. These invariants are used in the computations of 
the Born cross sections which appear in the MC cross sections expanded
to NLO, hence the dependence on the branching leg $L$ in eq.~(\ref{eq:dot}).
We also get
\begin{equation}
-2\bp_2\cdot\bk_2 = \bt_L+\Delm^2\,,\;\;\;\;\;\;
-2\bp_2\cdot\bk_1 = \bu_L-\Delm^2\,,
\end{equation}
where $\Delm^2=m_1^2-m_2^2$. As discussed in {\bf II}, the $\twototwo$ 
reduced invariants are functions of the invariants relevant to the 
$\twotothree$ kinematics. The computations required to determine 
such functions are non-trivial; we only report the results here:
\begin{eqnarray}
&&\bs_\pm = s+v_1+v_2\,,
\label{eq:bs}
\\
&&\bs_{\fot} = s\,,
\label{eq:bso}
\\
&&\bt_\pm = -\frac{\bs_\pm}{2}\left[1-\frac{x_2(t_1-u_1)+x_1(t_2-u_2)}
{2s\sqrt{x_+^2 -x_1x_2v_1v_2/s^2}}\right]
-\frac{\Delm^2}{2}\left(1+\frac{x_-}{\sqrt{x_+^2 -x_1x_2v_1v_2/s^2}}\right),
\nonumber\\*&&
\label{btfpm}
\\
&&\bt_{\fo}= -\half s\left[1-\left(\frac{t_2-u_1}{s-w_1}\right)
\frac{\bbeta}{\beta_2}\right]-\half\Delm^2\,,
\label{btfo}
\\
&&\bt_{\ft}= -\half s\left[1-\left(\frac{t_1-u_2}{s-w_2}\right)
\frac{\bbeta}{\beta_1}\right]-\half\Delm^2\,,
\label{btft}
\\
&&\bu_\pm = -\bs_\pm-\bt_\pm\,,\;\;\;\;\;\;\;\;
\bu_{\fot}=-s-\bt_{\fot}\,,
\label{eq:bu}
\end{eqnarray}
where
\begin{eqnarray}
\bbeta&=&\sqrt{1-2\frac{\Sigm^2}{\bs}+\frac{\Delmf}{\bs^2}}\,\,,
\label{bbetadef}
\\
\beta_1&=&\sqrt{\left(1+\frac{\Delm^2}{s-w_2}\right)^2-
              \frac{4sm_1^2}{(s-w_2)^2}}\,\,,
\\
\beta_2&=&\sqrt{\left(1-\frac{\Delm^2}{s-w_1}\right)^2-
              \frac{4sm_2^2}{(s-w_1)^2}}\,\,,
\label{eq:beta2}
\end{eqnarray}
and $\Sigm^2=m_1^2+m_2^2$. The $\twotothree$ invariants that appear
on the r.h.s. of eqs.~(\ref{eq:bs})--(\ref{eq:bu}) are labelled as
in {\bf II}; their definitions are also reported here in table~\ref{tab:kin}.
Equations~(\ref{eq:dot})--(\ref{eq:beta2}) give sufficient information, 
with tables~\ref{tab:sch}--\ref{tab:tcht}, to compute the shower scales 
to be used in eqs.~(\ref{eq:shin}) and~(\ref{eq:shout}).
\begin{table}[htb]
\begin{center}
\begin{tabular}{|c|c|c|}
\hline
Label & Invariant & Relation\\
\hline\hline
$s$   &  $2 p_1\cdot p_2$ & \\
$t_1$ & $-2 p_1\cdot k_1$ & \\
$t_2$ & $-2 p_2\cdot k_2$ & \\
$u_1$ & $-2 p_1\cdot k_2$ & \\
$u_2$ & $-2 p_2\cdot k_1$ & \\
$v_1$ & $-2 p_1\cdot k_3$ & $-s-t_1-u_1$\\
$v_2$ & $-2 p_2\cdot k_3$ & $-s-t_2-u_2$\\
$w_1$ &  $2 k_1\cdot k_3$ & $s+t_2+u_1-m_1^2+m_2^2$\\
$w_2$ &  $2 k_2\cdot k_3$ & $s+t_1+u_2+m_1^2-m_2^2$\\
$M^2_{12}$ & $(k_1+k_2)^2 $ &  $s+v_1+v_2$\\
\hline
\end{tabular}
\caption{\label{tab:kin}Notation for $\twotothree$ kinematics.}
\end{center}
\end{table}

We finally summarize the formulae for \HW\ showering variables.
The case of initial-state emissions is identical to that studied in {\bf II}, 
the condition $m_1\ne m_2$ being irrelevant here. When parton 1 branches, 
the showering variables $z_+$ and $\xi_+$ are related to the invariants 
as in \FNWeq{4.31} and \FNWeq{4.32}:
\begin{eqnarray}
v_1&=&-2\frac{1-z_+}{z_+^2}\xi_+\,E_0^2\,,
\label{ivz1}
\\
-\frac{v_2}{s}&=&\half (1-z_+)(2-\xi_+)\;.
\label{ivz2}
\end{eqnarray}
Using eq.~(\ref{shwrscale}) we can write the solutions explicitly:
\begin{eqnarray}
z^{(l)}_+ &=& \frac{2|\bl|}{v_1}\left[1-\sqrt{1-\frac{v_1}{|\bl|}
\left(1+\frac{v_2}{s}\right)}\,\right]\,,
\label{eq:z_ini}
\\ 
\xi^{(l)}_+ &=& 2\left[1+\frac{v_2}{s(1-z^{(l)}_+)}\right]\,,
\label{eq:xi_ini}
\end{eqnarray}
which are identical to \FNWeq{4.33} and \FNWeq{4.34} except for the 
different definition of the scale $\bl$.

The branching of parton 2 will be described in terms of the variables $z_-$ 
and $\xi_-$; these can be obtained from eqs.~(\ref{ivz1})--(\ref{eq:xi_ini})
by interchanging variables $v_1$ and $v_2$.

The formulae for final-state emissions are affected by the condition
$m_1\ne m_2$. When the parton with momentum $k_1$ branches, \FNWeq{4.23}
and \FNWeq{4.24} still formally hold
\begin{eqnarray}
w_1&=&2z_{\fo}(1-z_{\fo})\xi_{\fo} E_0^2\;,
\\
\zeta_{\fo}&=&(1-z_{\fo})\frac{1+(1-z_{\fo}\xi_{\fo})/\tbeta_1}{1+\tbeta_1}\,,
\end{eqnarray}
with
\begin{eqnarray}
\tbeta_1&=&\sqrt{1-(w_1+m_1^2)/E_0^2}\,,
\label{ivz3}
\\
\zeta_{\fo}&=&\frac{(2s-(s-w_1)\vep_2)w_2+
(s-w_1)\left[(w_1+w_2)\beta_2-\vep_2 w_1\right]}
{(s-w_1)\beta_2\left[2s-(s-w_1)\vep_2+(s-w_1)\beta_2\right]}\,,
\label{zetanew}
\\
\vep_2&=&1-\frac{\Delm^2}{s-w_1}\,.
\end{eqnarray}
It is apparent that eq.~(\ref{zetanew}) coincides with \FNWeq{4.27}
when $m_1=m_2$ (i.e. $\vep_2=1$). Solving eqs.~(\ref{ivz3}) 
and~(\ref{zetanew}) we obtain
\begin{eqnarray}
z^{(l)}_{\fo} &=& 1-\tbeta_1\zeta_{\fo} -\frac{w_1}{2(1+\tbeta_1)|\bl|}\,,
\label{eq:z_fin}
\\
\xi^{(l)}_{\fo} &=& \frac{w_1}{2z^{(l)}_{\fo}(1-z^{(l)}_{\fo})|\bl|}\,,
\label{eq:xi_fin}
\end{eqnarray}
which are identical to \FNWeq{4.28} and \FNWeq{4.29} except for the 
different definition of the scale $\bl$.

The branching of the parton with momentum $k_2$ can be treated along the
same lines. The showering variables $z^{(l)}_{\ft}$ and $\xi^{(l)}_{\ft}$
will be obtained from eqs.~(\ref{eq:z_fin}) and~(\ref{eq:xi_fin}) by
formally interchanging labels 1 and 2. Note that in this way the
quantity $\vep_1$ appears in the expression for $\zeta_{\ft}$, and
\begin{equation}
\vep_1=1+\frac{\Delm^2}{s-w_2}\,.
\end{equation}

\section{MC subtraction terms\label{sec:MCsubt}}
In this section, we construct explicitly the MC subtraction terms for
single-$t$ production, expressing them in terms of the variables used 
in the NLO computation. In order to do this, we start by introducing 
the phase-space parametrizations used in ref.~\cite{Frixione:1995ms} 
to deal with initial- and final-state emissions; in both cases, we 
integrate out the trivial azimuthal angles. We have
\begin{eqnarray}
d\phi_3^{(\INm)}&=&\frac{s}{1024\pi^4}\,\bbeta\left((1-\xi_i)s\right)\,
\xi_i\, d\xi_i dy_i d\cos\theta d\varphi\,,
\label{phspin}
\\
d\phi_3^{(\OUTm)}&=&\frac{s}{512\pi^4}\,\frac{\xi_j}{2-\xi_i(1-y_j)}\,
\xi_i\, d\xi_i dy_i dy_j d\varphi_j\,,
\label{phspout}
\end{eqnarray}
where $\bbeta(s)$ is given in eq.~(\ref{bbetadef}), and\footnote{Since
this section is specific to single-$t$ production, we set $m_2=0$ here.}
\begin{equation}
\xi_j=\frac{2(1-m_1^2/s-\xi_i)}{2-\xi_i(1-y_j)}\,.
\end{equation}
The variables labelled with index $i$ refer to the FKS parton (see
\FKSeq{4.3}), and those labelled with index $j$ refer to the massless
final-state parton that can become collinear to the FKS parton (see
\FKSeq{4.57}).  Note that $\xi_i$ is related to the variable $x$ used 
in {\bf I} and {\bf II} by the following equation
\begin{equation}
x\equiv 1-\xi_i\,.
\end{equation}
This implies that eq.~(\ref{phspin}) coincides with \FNWeq{B.22}.
We rewrite the real-emission finite contributions to the single-$t$ cross 
section (\FKSeq{4.37} and \FKSeq{4.65}) as follows:
\begin{eqnarray}
d\sigma_i^{(in,f)}&=&\half\xic\left[\omyid+\opyid\right]
\left((1-y_i^2)\xi_i^2\matr\right)\Szi\, d\tilde{\phi}_3^{(\INm)}\,,
\phantom{a}
\label{FKSrin}
\\
d\sigma_{ij}^{(out,f)}&=&\xic\omyjd
\left((1-y_j)\xi_i^2\matr\right)\Soij\, d\tilde{\phi}_3^{(\OUTm)}\,,
\label{FKSrout}
\end{eqnarray}
where 
\begin{equation}
d\phi_3^{(\INm)}=\xi_i\, d\tilde{\phi}_3^{(\INm)}\,,\;\;\;\;\;\;\;\;
d\phi_3^{(\OUTm)}=\xi_i\, d\tilde{\phi}_3^{(\OUTm)}\,.
\end{equation}
As discussed in {\bf I} and {\bf II}, MC subtraction terms 
can be obtained from the MC cross sections expanded to NLO.
Thus, following \FNWeq{B.21}, in order to construct them we must
write eq.~(\ref{eq:shin}) in the same form as eq.~(\ref{FKSrin}) 
(after relabeling), and eq.~(\ref{eq:shout}) in the same form as 
eq.~(\ref{FKSrout}) (after relabeling).
In order to do this, we note that the Born cross sections that
appear in the MC subtraction terms have the following forms (in order
to simplify the notation, we neglect here most of the indices)
\begin{eqnarray}
\dsb&=&\matb(\bs_\pm,\bt_\pm)\,\frac{\bbeta(\bs_\pm)}{16\pi}\,
d\cos\theta_{in}\,,
\\
\dsb&=&\matb(\bs_{\fsa},\bt_{\fsa})\,\frac{\bbeta(\bs_{\fsa})}{16\pi}\,
d\cos\theta_{out}\,,
\end{eqnarray}
for initial- and final-state branchings respectively. Here $\matb$ 
is the Born matrix element, and the angles $\theta_{in}$ and
$\theta_{out}$ have been introduced in \FNWeq{B.32} and \FNWeq{B.33}
respectively. As discussed in {\bf II}, it is not restrictive to obtain 
these scattering angles in the zero-angle-emission limits, which leads to
\begin{equation}
\theta_{in}=\theta\,,\;\;\;\;\;\;\;\;
\theta_{out}=y_j\,,
\label{thetas}
\end{equation}
where $\theta$ and $y_j$ are integration variables in eqs.~(\ref{phspin})
and~(\ref{phspout}) respectively. The first relation in eq.~(\ref{thetas})
coincides with \FNWeq{B.35}. We also note that, in the zero-angle-emission
limits, the (trivial) azimuthal angles generated by the showers can be
chosen to coincide with the angles $\varphi$ and $\varphi_j$ of 
eqs.~(\ref{phspin}) and~(\ref{phspout}) for initial- and final-state 
branchings respectively. We shall therefore insert the factors 
$d\varphi/(2\pi)$ and $d\varphi_j/(2\pi)$ in eqs.~(\ref{eq:shin}) 
and~(\ref{eq:shout}). We rewrite eq.~(\ref{eq:shin}) as follows
\begin{eqnarray}
d\hat\sigma^{(\pm)}\xMCB&=&
\frac{1}{(1-y_i^2)\xi_i}\,(1-y_i^2)\xi_i\,
\frac{\gs^2}{256\pi^4}\,\frac{P(z_\pm)}{\xi_\pm}\,
\frac{\partial(\xi_\pm,z_\pm)}{\partial(\xi_i,y_i)}
\nonumber \\&\times&
\matb(\bs_\pm,\bt_\pm)\,\stepf\left((z_\pm^{(l)})^2-\xi_\pm^{(l)}\right)
\bbeta(\bs_\pm)\,d\xi_i dy_i d\cos\theta d\varphi\,,
\end{eqnarray}
where the first factor on the r.h.s. matches the ``event'' part of 
eq.~(\ref{FKSrin}), i.e. that obtained by replacing the distributions 
with ordinary functions. Using eqs.~(\ref{phspin}) and~(\ref{eq:bs}), 
we get
\begin{equation}
\bbeta(\bs_\pm) d\xi_i dy_i d\cos\theta d\varphi=
\frac{1024\pi^4}{s}d\tilde{\phi}_3^{(\INm)}\,.
\end{equation}
Therefore
\begin{eqnarray}
d\hat\sigma^{(\pm)}\xMCB&=&\frac{1}{2\xi_i}
\left[\frac{1}{1-y_i}+\frac{1}{1+y_i}\right]
\left((1-y_i^2)\xi_i^2
\frac{d\Sigma^{(L=\pm)}}{d\phi_3^{(\INm)}}\xMCBB\right)
d\tilde{\phi}_3^{(\INm)}\,,
\label{MCcntin}
\\
\frac{d\Sigma^{(L=\pm)}}{d\phi_3^{(\INm)}}\xMCBB&=&\frac{4\gs^2}{s}\,
\frac{P(z_\pm)}{\xi_i\xi_\pm}\,
\frac{\partial(\xi_\pm,z_\pm)}{\partial(\xi_i,y_i)}\,
\matb(\bs_\pm,\bt_\pm)\,
\stepf\left((z_\pm^{(l)})^2-\xi_\pm^{(l)}\right),
\label{MCinfun}
\end{eqnarray}
which is identical (up to notational differences) to \FWeq{A.72} and
\FWeq{A.73}. The MC subtraction terms that enter eq.~(\ref{eq:rMCatNLO})
are readily obtained from eq.~(\ref{MCinfun}) using \FWeq{4.18}:
\begin{equation}
\frac{d\bSigma^{(L=\pm)}}{d\phi_3^{(\INm)}}\xMCBB=
\frac{\partial(\bx_{1i},\bx_{2i})}{\partial(x_1,x_2)}
\frac{d\Sigma^{(L=\pm)}}{d\phi_3^{(\INm)}}\xMCBB\,.
\label{SigtobSig}
\end{equation}
The reduced Bjorken $x$'s $\bx_{1i}$ and $\bx_{2i}$ are given in
\FNWeq{4.20} or \FNWeq{4.22}.

We now turn to the case of final-state emissions, and we rewrite 
eq.~(\ref{eq:shout}) as follows
\begin{eqnarray}
d\hat\sigma^{(\fsa)}\xMCB&=&
\frac{1}{(1-y_j)\xi_i}\,(1-y_j)\xi_i\,
\frac{\gs^2}{256\pi^4}\,\frac{P(z_{\fsa})}{\xi_{\fsa}}\,
\frac{\partial(\xi_{\fsa},z_{\fsa})}{\partial(\xi_i,y_j)}\,
\matb(\bs_{\fsa},\bt_{\fsa})
\nonumber \\&\times&
\stepf\left(1-\xi_{\fsa}^{(l)}\right)
\stepf\left(z_{\fsa}^{(l)}-\frac{m_\alpha}{E_0\sqrt{\xi_{\fsa}^{(l)}}}\right)
\bbeta(\bs_{\fsa})\,d\xi_i dy_i dy_j d\varphi_j\,.
\label{outtmp}
\end{eqnarray}
Using eq.~(\ref{phspout}) we get
\begin{equation}
d\xi_i dy_i dy_j d\varphi_j=
\frac{512\pi^4}{s}\,\frac{2-\xi_i(1-y_j)}{\xi_j}\,
d\tilde{\phi}_3^{(\OUTm)}\,.
\end{equation}
Inserting this equation into eq.~(\ref{outtmp}), and using eq.~(\ref{eq:bso})
we obtain
\begin{eqnarray}
d\hat\sigma^{(\fsa)}\xMCB&=&\frac{1}{\xi_i}
\frac{1}{1-y_j}\,
\left((1-y_j)\xi_i^2\frac{d\bSigma^{(L=f_\alpha)}}{d\phi_3^{(\OUTm)}}\right)
d\tilde{\phi}_3^{(\OUTm)}\,,
\label{MCcntout}
\\
\frac{d\bSigma^{(L=f_\alpha)}}{d\phi_3^{(\OUTm)}}&=&
\frac{2\gs^2}{s}\,\frac{2-\xi_i(1-y_j)}{\xi_i\xi_j}\,
\bbeta(s)\,
\frac{P(z_{\fsa})}{\xi_{\fsa}}\,
\frac{\partial(\xi_{\fsa},z_{\fsa})}{\partial(\xi_i,y_j)}\,
\matb(s,\bt_{\fsa})
\nonumber \\*&\times&
\stepf\left(1-\xi_{\fsa}^{(l)}\right)
\stepf\left(z_{\fsa}^{(l)}-\frac{m_\alpha}{E_0\sqrt{\xi_{\fsa}^{(l)}}}\right).
\label{MCoutfun}
\end{eqnarray}
Note that we directly defined $\bSigma$ (rather than $\Sigma$ as in
eq.~(\ref{MCcntin})) thanks to \FNWeq{4.7}.

We have checked analytically that the MC counterterms introduced above
locally cancel the collinear divergences of the real matrix elements.
As already discussed in {\bf I} and {\bf II}, this happens in the soft 
limit only after angular integration. We therefore adopt here the same 
solutions as in \FNWeq{B.43}. As in the previous cases, we checked that 
the parametric dependence introduced in this way is totally negligible, 
as we expect from power-suppressed effects.


\providecommand{\href}[2]{#2}\begingroup\raggedright\endgroup

\end{document}